\documentclass[preprint,3p,twocolumn]{elsarticle}

\usepackage{graphicx}
\usepackage{subfigure}
\usepackage{color}
\usepackage{placeins} % FloatBarrier 
\usepackage{amsfonts,amsmath,amssymb}
\usepackage{breqn}
\usepackage{multirow}
\usepackage{braket}
\usepackage{numcompress}
\makeatletter
\def\ps@pprintTitle{%
 \let\@oddhead\@empty
 \let\@evenhead\@empty
 \def\@oddfoot{}%
 \let\@evenfoot\@oddfoot}
\makeatother

\begin{document}

\begin{frontmatter}

\title{Mechanical, elastic and thermodynamic properties of crystalline lithium silicides}
%\tnotetext[mytitlenote]{Fully documented templates are available in the elsarticle package on %\href{http://www.ctan.org/tex-archive/macros/latex/contrib/elsarticle}{CTAN}.}

\author[theophys]{Sebastian Schwalbe\corref{mycorrespondingauthor}} 
\cortext[mycorrespondingauthor]{Corresponding author}
\ead{schwalbe@physik.tu-freiberg.de}

\author[theophys]{Thomas Gruber}
\author[theochem]{Kai Trepte}
\author[physchem]{Franziska Biedermann} 
\author[physchem]{Florian Mertens}
\author[theophys]{Jens Kortus}
\address[theophys]{Institute of Theoretical Physics, TU Bergakademie Freiberg, Leipziger Str. 23, D-09596 Freiberg, Germany}
\address[theochem]{Theoretical Chemistry, Technische~Universit\"at~Dresden, Bergstra{\ss}e~66c, D-01062 Dresden, Germany}
\address[physchem]{Institute of Physical Chemistry, TU Bergakademie Freiberg, Leipziger Str. 29, D-09599 Freiberg, Germany}

\begin{abstract}
We investigate crystalline thermodynamic stable lithium silicides phases 
(Li$_{x}$Si$_{y}$) with density functional theory (DFT) and a force-field method based on modified embedded atoms (MEAM) and compare our results with experimental data. 
This work presents a fast and accurate framework to calculate thermodynamic properties of crystal structures with large 
unit cells with MEAM based on molecular dynamics (MD). 
Mechanical properties like the bulk modulus and the elastic constants are evaluated in addition to thermodynamic properties including the phonon density of states, the vibrational free energy and the isochoric/isobaric specific heat capacity for Li, Li$_{12}$Si$_{7}$, Li$_{7}$Si$_{3}$, Li$_{13}$Si$_{4}$, Li$_{15}$Si$_{4}$, Li$_{21}$Si$_{5}$, Li$_{17}$Si$_{4}$, Li$_{22}$Si$_{5}$ and Si.
For a selected phase (Li$_{13}$Si$_{4}$) we study the effect of a temperature dependent phonon density of states and its effect on the isobaric heat capacity. 
\end{abstract}


\end{frontmatter}

%\linenumbers

\section{Introduction}
The rapid growth of new technologies (e.g. portable devices like smart-phones and tablets, electro-mobility or hybrid-cars etc.) demands new and efficient energy storage systems. The leading graphite anode material is limited due to a low Li storage capacity $C_{\text{s}} \approx$ 372 mAh g$^{-1}$ \cite{LiSi_exp_ref5}.
The development of a new improved energy storage system requires a material which has a higher specific energy density, current density and charge/discharge cycling stability. To address these challenges a thorough scientific understanding of these materials is required.\\
Lithium silicides (Li$_{x}$Si$_{y}$) are discussed as a new kind of lithium ion battery materials due to their high lithium storage capacity ($C_{\text{s}} \approx$ 4200 mAh g$^{-1}$ for Li$_{22}$Si$_{5}$) \cite{LiSi_exp_ref6}.
Several experimental and ab-initio investigations were carried out in the last years to determine the specific heat capacity and related properties to construct the phase diagram of Li$_{x}$Si$_{y}$ \cite{LiSiZintl3,LiSi_exp,LiSi_MEAM,Li13Si4}. 
On the one hand experiments need highly accurate sample preparation and measurements and with that an appropriate time span for the evaluation of thermodynamic data. 
On the other hand some of the Li$_{x}$Si$_{y}$ phases have large unit cells with hundreds of atoms, resulting in a high computational effort to describe such systems theoretically. \\
\citet{LiSi_MEAM} started to create Li$_{x}$Si$_{y}$ modified embedded atom potentials and calculated elastic and dynamic properties of Li$_{x}$Si$_{y}$ from amorphous phases. In this work we focus on the crystalline structures
of the thermodynamically stable Li$_{x}$Si$_{y}$ phases and calculate elastic properties like the bulk modulus and elastic constants as well as thermodynamic properties like the vibrational free energy and phonon density of states in the framework of density functional theory and the force-field method of modified embedded atoms. We show how to overcome the limit of the quasi-(harmonic) approximation and present a temperature dependent phonon density of states calculated from molecular dynamics simulations. All our computational results are compared to available experimental data \cite{LiSi_exp}.   
\section{Theory - Computational methods}

One important physical property for characterizing lithium ion batteries is the so-called specific storage capacity $C_{\text{s}}$. 
For lithium silicides Li$_{x}$Si$_{y}$ the specific storage capacity $C_{\text{s}}$ can be calculated using
\begin{equation}
    C_{\text{s}} = x \cdot F/(y \cdot m_{\text{Si}}), 
\end{equation}
where $F = 96485.33$ C mol$^{-1}$ is the Faraday constant and $m_{\text{Si}} = 28.08$ g mol$^{-1}$ is the molar mass of silicon. 

\begin{table}[h!]
\setlength{\tabcolsep}{1.5pt}
\tiny
\centering
\caption{Theoretical specific lithium storage capacity $C_{\text{s}}$ of Li$_{x}$Si$_{y}$ crystal structures.\newline} 
\centering
 \begin{tabular}{l|c|c}
 crystal structure & Li mass content [\%] & $C_{\text{s}}$ [mAh g$^{-1}$]\\ \hline
 Si & 0 & 0\\
 LiSi & 19.8 & 954\\
 Li$_{12}$Si$_{7}$ & 29.8 & 1636\\
 Li$_{7}$Si$_{3}$  & 36.6 & 2227 \\
 Li$_{13}$Si$_{4}$ & 44.5 & 3102 \\
 Li$_{15}$Si$_{4}$ & 48.1 & 3579\\
 Li$_{21}$Si$_{5}$ & 50.9 & 4009 \\
 Li$_{17}$Si$_{4}$ & 51.2 & 4056\\
 Li$_{22}$Si$_{5}$ & 52.1 & 4200  \\
 Li & 100 & --- \\
 \end{tabular}
\label{storage_capacity}
\end{table}

The understanding of thermodynamic material properties can be based on three pillars: experiment, atomistic simulations, and so-called CALPHAD modelling \cite{CALPHAD_summary}. Experimental methods (XRD, neutron-scattering, differential scanning calorimetry etc.) can deliver structural information about battery materials, 
but often even neutron-scattering is not able to determine lithium position in crystal structures accurately. 
Many physical properties (e.g. phonon densities, free energies, heat capacities etc.) depend on the crystal structure. Atomistic electronic structure methods can often clarify structural uncertainties with the help of energy surface calculations, relaxation methods, or molecular dynamics. Computational methods can also be employed to calculate thermodynamic data from an experiment independent point of view \cite{Li13Si4}.\\
This data is useful to get insight in complex physical behavior like
diffusion effects \cite{Li13Si4} or phase transitions \cite{Steve}. Additionally, theory at these levels can provide thermodynamic data for structures which are not accessible by experiments (e.g metastable phases). \\
The CALPHAD method can be based on input from both experiment as well as atomistic calculations to model phase diagrams for binary, ternary or even more complex systems.
In this work we concentrate on computational aspects of thermodynamics (also called computational thermodynamics). 
Important physical properties are calculated from structural input quantities without any fit to experimental data.
The field of applicable ab-initio theories ranges from highly accurate quantum chemistry methods (e.g. Hartree Fock, configuration interaction, coupled cluster methods etc.) to more approximate methods (e.g density functional theory, force-field methods, phase field methods etc.) 
often used in materials research. 
All computational methods are based on approximations and therefore all of these methods have their own limits. 
State-of-the-art computational thermodynamics is based on DFT, which is limited by the large computational cost. The advantage of this method is the obtained accuracy for total energies and forces.
Thermodynamics based on molecular dynamics is limited by the given accuracy of the used force-field approximation. The advantage of such methods is that configuration changes, diffusion effects or very large volume expansions or contractions can be calculated in an appropriate computational time. Additionally, force-field methods offer the opportunity for long time molecular dynamics and thus reliable statistics.
%The ergodic theorem enables the possibility to treat configuration effects (e.g. diffusion) in a more accurate manner.
%while losing accuracy in the resulting total energies and forces for specific configurations. 
In this work we show a pathway to combine the advantages of both methods to get a deeper understanding of 
the thermodynamics of lithium silicides. 
As mentioned before we concentrate on the crystalline structures of Li$_{x}$Si$_{y}$. In this respect density functional theory (DFT) \cite{DFT_Martin} calculations are very accurate and deliver a good description of the electronic structures of such systems. 
The framework of DFT is used to calculate the ground state electron density $n(\textbf{r})$ of a system. From this quantity one can determine the total energy $E_{\text{tot}}[n(\textbf{r})]$ at $T$ = 0 K. Additionally, the forces between atoms can be evaluated using the Hellmann-Feynman theorem \cite{DFT_Martin,DFT_FORCE_STEVE} and even the stress tensor can be determined \cite{DFT_STRESS_STEVE1,DFT_STRESS_STEVE2}.
One of the most important concepts is the Born-Oppenheimer approximation, which is implicitly assumed. This approximation states that electronic degrees of freedom vary much faster than nuclear ones, i.e. electrons follow the motion of the nuclei almost instantaneously. 

This fact allows the separation of electron and nuclear dynamics, while the nuclear motion can be treated classically. Using this concept enables the calculation of solid state (and other) properties with DFT by only solving the electronic problem.

In contrast to the force-field method the computational effort of DFT highly depends on the number of atoms in the unit cell.
Some of the Li$_{x}$Si$_{y}$ phases have crystal structures with many atoms (see TAB. \ref{structures}) and ab-initio thermodynamics becomes very expensive. 
For that reason we additionally used the force-field based molecular dynamics method to calculate thermodynamic properties for all known
phases of the Li$_{x}$Si$_{y}$ system. We used the so-called modified embedded atom method (MEAM), in which
the total energy $E_{\text{tot,MEAM}}$ of a system of atoms can be written in this formalism \cite{LAMMPS_MEAM,Alam} as
\begin{align}
 E_{\text{tot,MEAM}} = \sum_i E_{\text{tot,MEAM},i},
\end{align}
where the energy of the $i$-th atom is defined as
\begin{align}
 E_{\text{tot,MEAM},i} = \underbrace{E_{\text{EB},i}(\textbf{n}_{\text{MEAM},i})}_{\text{embedding energy}} + \underbrace{\frac{1}{2} \sum_{i \neq j} \phi_{ij}(\textbf{r}_{ij})}_{\text{pair potential term}}.
\end{align}
The embedding energy $E_{\text{EB},i}$ is a function of the background electron density $\textbf{n}_{\text{MEAM},i}$ at the site of atom $i$ whereas the pair potential $\phi_{ij}(\textbf{r}_{ij})$ between
atoms $i$ and $j$ depends on the distance $\textbf{r}_{ij}$. The embedding energy $E_{\text{EB},i}$ is the energy which is needed to insert an atom $i$ at a site
where the background electron density equals $\textbf{n}_{\text{MEAM},i}$. A detailed description of the method itself can be found in \citet{LAMMPS_MEAM}.
\section{Elastic and thermodynamic properties} 
Hook's law connects the stresses $\sigma_{ij}$ (stress tensor) and strains $\varepsilon_{ij}$ (strain tensor) inside a material by the following linear relationship 
\begin{equation}
 \sigma_{ij} = \sum_{k=l}^{3}\sum_{l=1}^{3} c_{ijkl} \varepsilon_{kl},
\end{equation}
where $c_{ijkl}$ are elastic constants of the material. 
This linear relation holds in the limit of infinitesimal deformation. 
% Bulk modulus 
The bulk modulus $B$ is an important material property. 
A general expression for the bulk modulus $B_{\text{Voigt}}$ depending on the elastic constants $c_{ij}$ was given by Voigt \cite{B_Voigt} 
\begin{equation}
  B_{\text{Voigt}} = \frac{(c_{11} + c_{22} + c_{33})+2 (c_{12} + c_{23} + c_{13})}{9}.
  \label{B_Voigt}
\end{equation}
Another possibility for the numerical determination of the bulk modulus is given by the energy-volume relation of the system 
\begin{equation}
 B_{\text{EOS}}(V) = V \biggl(\frac{\partial^2 E(V)}{\partial V^2}\biggr)_{T,S} \label{B_EOS},
\end{equation}
where the volume dependent energy $E(V)$ can be calculated using an equation of state (EOS), like the Birch-Murnaghan EOS \cite{EOS}.\\
% Lattice dynamics Phonons 
Lattice dynamics is a broad field and the most commonly used methods are based on the harmonic and the quasi-harmonic approximation (QHA). 
The harmonic approximation (HA) in the field of DFT is used for example to calculate IR- and RAMAN spectra for molecular systems while 
the framework of the quasi-harmonic approximation is generally used to calculate thermodynamic properties of bulk systems. There are two common methods in DFT, the direct method (also called frozen phonon method) and the density functional perturbation theory (DFPT, also called linear response).
Both methods allow to calculate the dynamical matrix. 
This dynamical matrix can be used to calculate the phonon dispersion relation or the phonon density of states (PDOS). 
In contrast to DFPT in which only unit cells are used the direct method is based on supercells. 
%For the direct method supercells are needed, while in DFPT one only needs the unit cell. 
Thus, the direct method is computationally more demanding than DFPT. However, it is possible to extend the supercell approach to include anharmonic effects \cite{supercell_anharmonic}, 
while DFPT is restricted to harmonic effects. For large unit cells as in the case of Li$_{x}$Si$_{y}$ phases both of these methods become 
computationally demanding. In the field of classical molecular dynamics there are two methods to calculate the phonon density of states as well. 
Both methods are based on the fluctuation dissipation theorem. The first one uses the velocity autocorrelation function $C_{\text{VAF}}(t) = \braket{\textbf{v}(0)\textbf{v}(t)}/\braket{\textbf{v}(0)\textbf{v}(0)}$ to calculate the vibrational density of states (VDOS, also called PDOS or power spectra) from the velocities $\textbf{v}$ of the atoms (during a long MD run) with 
\begin{equation}
 \text{PDOS}(\omega) = \biggl|\frac{1}{\sqrt{2\pi}} \int_{-\infty}^{\infty} dt e^{i\omega t} C_{\text{VAF}}(t) \biggr |^2. 
\end{equation}
% isobaric heat capacity
The isochoric heat capacity can be calculated from this vibrational/phonon density of states PDOS$(\omega)$ using
\begin{align}
 C_{V}(T) &= \biggl(\frac{\partial E_{\text{vib}}}{\partial T}\biggr)_{V} \\
	  &= k_{\text{B}} \int \text{PDOS}(\omega) \biggl(\frac{\hbar \omega}{2 k_{\text{B}} T}\biggr)^2 \frac{d \omega}{\sinh^2{\frac{\hbar \omega}{2 k_{\text{B}} T}}}.
\end{align}
The second method is based on the fluctuations/displacements of atoms during a molecular dynamics simulation \cite{GreensFunction,fix_phonon,fix_phonon2}.
During the MD run the displacements of the atoms are recorded. %are observed.
Afterwards Green's function coefficients are evaluated in real space, which are given by the ensemble average of the scalar product of the atomic displacements \cite{GreensFunction}.
From these Green's function coefficients the force constants, the dynamical matrix, and with that PDOS$(\omega)$ can be calculated. 
% linear thermal expansion coeffient 
The linear thermal expansion coefficient $\alpha_{l}$ can be determined according to
\begin{equation}
 \alpha_{l} = \frac{1}{V}\biggl(\frac{\partial V}{\partial T}\biggr)_{p}.
\end{equation}
For the evaluation of this expression it is necessary to perform a set of $N$ molecular dynamics simulation with NPT ensembles
for different temperatures $T$ and fixed pressures $p$. From a specific MD run the mean volume $\braket{V}_{\text{NPT},r}$ and for all MD runs the global volume minimum $\min_{\forall r}{\braket{V}_{\text{NPT},r}}$
can be calculated. This leads to following expression of the linear thermal expansion coefficient 
\begin{equation}
  \alpha_{l} = \frac{1}{N} \sum_{r=1}^{N} \frac{\Delta \braket{V^{'}}_{\text{NPT},r}}{\Delta  \braket{T}_{\text{NPT},r}},
\end{equation}
with $r$ indicating a specific MD run. The reduced volume $\braket{V^{'}}_{\text{NPT},r}$ is given by 
\begin{equation}
  \braket{V^{'}}_{\text{NPT},r} =  \biggl (\frac{\braket{V}_{\text{NPT},r}}{\min_{\forall r}{\braket{V}_{\text{NPT},r}}}\biggr)_{p}.
\end{equation}

% isochoric heat capacity 
With the knowledge of the bulk modulus $B$, the linear thermal expansion coefficient $\alpha_{l}$ and the isochoric heat capacity $C_{V}$ we can calculate the isobaric heat capacity $C_{p}$ by using
\begin{equation}
 C_{p} - C_{V} = nVBT\alpha_{l}^2,
 \label{equation_cp}
\end{equation}
with $n$ being the mole number. 
% rpdf 
Sometimes the characteristics of structural changes in huge simulation cells during MD simulations are difficult to analyze. In these cases the so called radial pair distribution function (RPDF) $g(r)$ can deliver information on changes in the coordination sphere of specified atoms at different temperatures. 
The radial pair distribution function
%\begin{equation}
%    g(r) = \frac{V}{N^2} \braket{\sum\limits_{i\neq j = 1} \delta(r - |\textbf{R}_i - %\textbf{R}_j|)},
%\end{equation}
is the average number of atoms in a shell $[r,r+dr]$ around an atom at $r=0$. For each atom $i$ = 1, ..., $N$ the number of atoms $j$ in the shell with the radius $r_{ij} = |\textbf{R}_i - \textbf{R}_j|$ ($r < r_{ij} \le r+dr$) are counted \cite{rpdf}
\begin{equation}
    dN(r) = \rho 4 \pi r^2 g(r) dr = n(r) dr,
\end{equation}
where $\rho = N/V$ is the average density and the pair distribution function (PDF) $n(r)$ is given by 
\begin{equation}
    n(r) = \frac{1}{N} \braket{\sum\limits_{i\neq j = 1} \delta(r - |\textbf{R}_i - \textbf{R}_j|)}.
\end{equation}
Normalizing $dN(r)$ to the shell volume $dV(r) = 4\pi r^2 dr$
leads to the following expression for the RPDF,
\begin{align}
    g(r) &= \frac{P_1}{P_2}=\frac{dN(r)/dV(r)}{N/V} = \frac{n(r)dr}{\rho 4 \pi r^2 dr} \\
    g(r) &= \frac{V}{4 \pi r^2 N^2} \braket{\sum\limits_{i\neq j = 1} \delta(r - |\textbf{R}_i - \textbf{R}_j|)},
\end{align}
where $P_1$ is the probability to find an atom in a volume $dV(r)$ at a distance $r$ and $P_2$ is the probability 
to find an atom in a volume $dV(r)$ if the atoms where uniformly distributed \cite{rpdf2}.\\ 
% Anharmonics 
Besides the standard harmonic and quasi-harmonic approximations, newer developments and methods for an explicit inclusion of anharmonicity in ab-initio thermodynamics have been proposed by the group of Neugebauer, e.g. the upsampled thermodynamic integration using Langevin dynamics (UP-TILD) \cite{UP-TILD} approach. In this context a set of different molecular dynamics runs is used to construct a force-field, which reproduces the right temperature dependence. 
Furthermore, to mention some other important approaches here: there is one by the group of \citet{AnharmonicApproach}, the stochastic self-consistent harmonic approximation by \citet{AnharmonicApproach2}, and the temperature dependent effective potential (TDEP) method by \citet{AnharmonicApproach3} to include high-temperature anharmonicity. 
In the TDEP method, ab-initio molecular dynamics are used to fit the temperature dependent effective potential. All these approaches have in common that 
they try to represent the high-temperature anharmonicity by inclusion of the phonon-phonon coupling. 

Our MEAM based MD approach for the calculation does not include phonon-phonon coupling directly. In principle, however, the performed MEAM MD runs are fully anharmonic, thus higher order anharmonic effects may be projected onto the harmonic force constants. 
Further investigations to include the phonon-phonon coupling directly within our method are planned for future work. 

% our MD Cp method 

\begin{figure}[h!]
 \centering
 \includegraphics[width=0.45\textwidth]{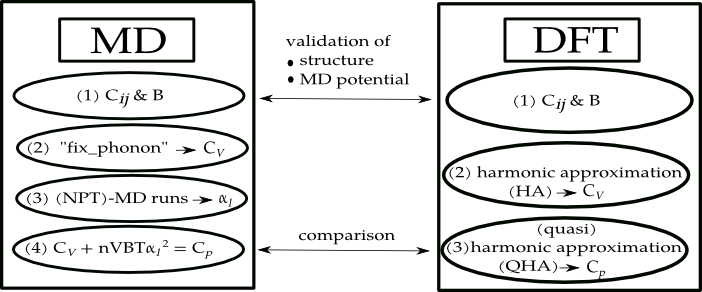}
 \caption{MD C$_{p}$ method as implemented in this work.}
 \label{md_cp}
\end{figure}
The following chapters are structured according to the procedure steps of our MD $C_{p}$ method (see FIG. \ref{md_cp}). 
First we will give a short summary of the codes and numerical parameters which we used in our calculations (see Section \ref{computational_details}). Then we discuss the first step of our MD $C_{p}$ method and present the elastic properties of the lithium silicides (see Section \ref{results1}). 
Within this step we show that the used MEAM potential is valid for the description of the lithium silicides crystal structures. In Section \ref{results2} we will present $C_{p}$ values for different 
lithium silicides calculated within our MD $C_{p}$ method and compare them with DFT and experiment. Additionally, we analyzed computational time aspects of the method in Section \ref{computational_time_analysis}. 

\section{Computational details\label{computational_details}}
The electronic structure calculations based on density functional theory were performed using the projector augmented
wave (PAW) method \cite{PAW} implemented in the QUANTUM ESPRESSO (QE) \cite{QE} code. The PAW potentials were created by means of the
atompaw code \cite{atompaw} with an exchange-correlation functional according to Perdew and Wang \cite{LDA}. For lithium, the 1s, 2s
and 2p states are projectors and thus no core states were used. The PAW sphere radii were set to 2.2 $a_{0}$. For silicon the 3s and
3p projectors were used; 1s2s2p are core states and the PAW sphere radii were set to 1.5 $a_{0}$.
The kinetic energy cutoff was adjusted for each Li$_{x}$Si$_{y}$ structure in the fashion to obtain a convergence better than
2 meV atom$^{-1}$. For specific cutoff values per phase see supplementary material TAB. A1. In addition, adjusted Monkhorst-Pack grids for the Brillouin zone integration for each phase (see supplementary material TAB. A1) were used to provide the same
k-point grid density in each direction. The self-consistency of total energy was set to 10$^{-8}$ Ry.
The phonon wave numbers are obtained from a fully relaxed unit cell. The variable cell optimization was carried
out with convergence thresholds in respect to forces of 10$^{-8}$ Ry $a^{-1}_{0}$ and pressures on the unit cell below 0.1 kbar.\\
The ab-initio elastic constants are calculated utilizing the ElaStic code \cite{ELASTIC} with QE. 
The vibrational wave numbers have been calculated using perturbation theory \cite{phonon_thomas} with a factor of two smaller q grid compared with the
k grid. For comparison, the same fully relaxed cell with the same convergence thresholds has been used for the finite
displacement method. The PHONOPY code \cite{Phonopy} was used to create 2x2x2 supercells with finite displacements and the
forces were then calculated with QE. A 3x3x3 supercell has been tested. A difference in the
free energy up to 3.6 meV at 1000 K and a difference in the heat capacity up to 0.16  JK$^{-1}$mol$^{-1}$ atom$^{-1}$ at 70 K has been found. Since 
the computational time for the 3x3x3 supercell is 30 times higher than for the 2x2x2 supercell and the difference are
small enough, the 2x2x2 supercell has been used in further calculations.\\
All force-field molecular dynamics calculations were performed in the LAMMPS code \cite{LAMMPS} by using
MEAM \cite{LAMMPS_MEAM}. As MEAM potential, we used the Li$_{x}$Si$_{y}$ optimized MEAM derived by \citet{LiSi_MEAM}.
All calculation using these parameters are denoted with M1. Only for pure silicon we also used the Si MEAM potential from the work of 
\citet{LiSi_MEAM}, which describes the Si-Si bond better than the Li$_{x}$Si$_{y}$ MEAM potential. 
We denoted these parameters for the bulk silicon MEAM calculations with M2. 
Starting from structural input from the ICSD, we used the PWTOOLS toolkit \cite{pwtools} and the latgen code \cite{latgen} to convert 
the cif file to LAMMPS structural input format. Also for the LAMMPS calculations we tested several supercell sizes for each calculated structure. The final supercell sizes are listed in supplementary material TAB. A2. 
We used the `fix\_phonon` \cite{fix_phonon,fix_phonon2} option to calculate the phonon density of states within the LAMMPS code. 
Each of these MD runs was carried out for 13 ns. For the calculation of the linear expansion coefficient we performed, depending on the actual structure, 15-20 NPT MD runs (see TAB. \ref{table_lammps_time}). The bulk modulus $B$ was calculated with the LAMMPS included ELASTIC module.
This module has been adjusted to calculate $B_{\text{Voigt}}$ (see Eq. (\ref{B_Voigt})). For the calculation of $B_{\text{EOS}}$ (see Eq. (\ref{B_EOS})) we performed 
with the LAMMPS code a set of thermostatic calculations with a conjugated gradient scheme and evaluated these data with the Birch-Murnaghan equation of state \cite{EOS}. 

\section{Results\label{results}}
Lithium silicides consist of small alkali-metal lithium atoms and bigger half-metal silicon atoms. The silicon atoms are not always isolated from each other. In fact, in some phases they tend to build rigid Si-Si clusters (e.g. dimers, rings etc.). 
There is an ongoing discussion whether or not lithium silicides can be classified with the well known Zintl concept \cite{Zintl} (see Refs. \citet{Li21Si5_icsd}, \citet{LiSiZintl2} and \citet{LiSiZintl3}). Zintl phases consist of one alkaline/alkaline earth metal and one element of the 3rd or 5th period. 
With that, Zintl phases show both ionic and covalent bonding parts. 
These conceptional aspects are important in respect to our goal to describe 
the class of lithium silicides with only one force-field. 
If some Li$_{x}$Si$_{y}$ phases may follow the Zintl concept and some may not, one 
force-field may not be sufficient to treat all Li$_{x}$Si$_{y}$ phases. \\
The structural parameters for each phase are given in TAB. \ref{structures}.
It should be noted that we transformed all phases in an orthogonal setup, which was necessary for the LAMMPS calculations.  
% discussion of the similarity of Li17Si4. Li21Si5, Li22Si5 
The Li$_{21}$Si$_{5}$, the Li$_{22}$Si$_{5}$ phase, and the Li$_{17}$Si$_{4}$ phase show a very similar lithium content (see TAB. \ref{structures})
and their lattice parameters are only slightly different. These structures are very hard to distinguish with experimental methods (e.g. X-ray diffraction or 
neutron-scattering). In case of Li$_{13}$Si$_{4}$ there exist diverse structural settings,
but we will exclusively focus on the structure found by \citet{Li7Si3_icsd} labeled for further investigations with Li$_{13}$Si$_{4}$ ICSD and the structure found by the evolution algorithm EVO \cite{EVO_CODE} by \citet{Li13Si4} labeled with Li$_{13}$Si$_{4}$ EVO. \\
In our simulations we can use the given structures and we are able to explicitly calculate each of these phases. This achievement enables us to provide 
further information regarding the stability of these phases.
\begin{table}[h!]
\setlength{\tabcolsep}{1.5pt}
\tiny
\centering
\caption{Crystal structure overview of all calculated Li$_{x}$Si$_{y}$ phases. The lattice parameters are given by $a$, $b$ and $c$ (all angles $\alpha$ = $\beta$ = $\gamma$ = $90^{\circ}$). In addition, SG is the spacegroup, $w_{\text{Li}}$ the Lithium mass content and $N_{\text{atoms}}$ the number of atoms per unit cell with and without symmetry.\newline} % is given in the last column
 \begin{tabular}{l|c|c|ccc|c}
 crystal structure & SG & $w_{\text{Li}}$ & $a$ & $b$ & $c$ & $N_{\text{atoms}} $ \\
 & & [\%] & [\AA] & [\AA] & [\AA] & \\ \hline 
 Si (Ref. \citep{Si_icsd}) & 227 & 0 & 5.43 & 5.43 & 5.43 & 1 (8)\\
% LiSi (Ref. \onlinecite{LiSi_icsd})  & 88 & 19.8 & 9.35 & 9.35 & 5.74 & 2 (32)\\
 Li$_{12}$Si$_{7}$ (Ref. \citep{Li12Si7_icsd}) & 62 & 29.8 & 8.60 & 19.76 & 14.34 & 22 (152) \\
 Li$_{7}$Si$_{3}$  (Ref. \citep{Li7Si3_icsd}, a)& 1 & 36.6 & 7.42 & 4.29 & 17.69 & 40 \\
 Li$_{13}$Si$_{4}$ EVO (Ref. \citep{Li13Si4}, a)& 55 & 44.5 & 7.75 & 14.56 & 4.34 & 9 (34) \\
 Li$_{13}$Si$_{4}$ ICSD (Ref. \citep{Li13Si4_icsd}) & 55& 44.5& 7.99& 15.21 & 4.43 & 9 (34) \\
 Li$_{15}$Si$_{4}$ (Ref. \citep{Li15Si4_icsd}) & 220 & 48.1 & 10.60 & 10.60 & 10.60 & 3 (76)\\
 Li$_{21}$Si$_{5}$ (Ref. \citep{Li21Si5_icsd})& 216 & 50.9 & 18.71 & 18.71 & 18.71 & 16 (416) \\
 Li$_{17}$Si$_{4}$ (Ref. \citep{Li21Si5_icsd}, b)& 216 & 51.2 & 18.71 & 18.71 & 18.71 & 17 (420)\\
 Li$_{22}$Si$_{5}$ (Ref. \citep{Li22Si5_icsd})& 216 & 52.1& 18.75 & 18.75 & 18.75 & 20 (432) \\
 Li (Ref. \citep{Li_icsd}) & 229 & 100 & 3.51 & 3.51 & 3.51 & 1 (2)\\
 \end{tabular}
\newline
\newline 
\tiny{a ...  Original cell is transformed in an orthogonal setting and values given for DFT optimized structure}
\newline
\tiny{b ...  Same cell as Li$_{21}$Si$_{5}$ with one additional Wykoff Li position}
\label{structures}
\vspace{-1.5em}
\end{table}
%\FloatBarrier
%\vfill
We have calculated the specific storage capacity $C_{\text{s}}$ for all Li$_{x}$Si$_{y}$ crystal structures (see TAB. \ref{storage_capacity}). 
With increasing lithium content, $C_{\text{s}}$ also increases from 1636 to 4200 mAh g$^{-1}$. 
In accordance with this observation, the phases with similar Li to Si ratio (Li$_{21}$Si$_{5}$, Li$_{22}$Si$_{5}$ and Li$_{17}$Si$_{4}$) only slightly differ in the respective absolute value. 

\subsection{Results I - elastic properties\label{results1}} 
For our MD $C_{p}$ method it is necessary to validate the force-fields which should be used for the computation of the isobaric heat capacity. 
The elastic constants describe the interplay of the resulting forces acting on different atoms by applying external deformation of the systems.
The right elastic behavior is the key property for any phonon calculation. We calculate the elastic properties for Li$_{x}$Si$_{y}$ phases with MEAM MD 
and compared them to our DFT results (see TAB. \ref{elastic_properties}).
In our MEAM MD calculations the energy volume relation was obtained using different structurally relaxed volumes. \\
The geometry optimization was performed with a conjugate gradient (cg) minimizer. 
We calculated the total energy for different volumes for each phase and used the resulting $E(V)$ to calculate $B_{\text{EOS}}$ with a Birch-Murnaghan equation of state. 
% Potential notes 
All phases besides the pure silicon are calculated with the same M1 MEAM potential. For silicon we used the M2 MEAM potential, because it describes the Si-Si bond better. \\
One remarkable result is that the obtained elastic constants of the crystalline Li$_{x}$Si$_{y}$ phases are 
in good agreement with those calculated by \citet{LiSi_MEAM} for the amorphous phases. 
In conclusion, this potential allows the description of the Li$_{x}$Si$_{y}$ system with amorphous and crystalline structures.   
% comparison between DFT and MEAM 
Generally, the DFT and M1/M2 values deliver similar values for each elastic property and the trends within the elastic constants are the same. 
% Li13Si4
The M1 calculations for the EVO Li$_{13}$Si$_{4}$ (M1 a, Ref. \citep{Li13Si4}) and ICSD Li$_{13}$Si$_{4}$ (M1 b, Ref. \citep{Li13Si4_icsd}) structures (see TAB. \ref{elastic_properties}) 
give the same magnitude and only slightly differ in the total value. 
% comparison Li17Si4. Li21Si5, Li22Si5 
Our DFT and M1 calculations show a similarity of the Li$_{21}$Si$_{5}$ and the Li$_{22}$Si$_{5}$ phase in all elastic properties, 
whereas the Li$_{17}$Si$_{4}$ phase differs more from these stoichiometrically similar phases. 
In summary, the used MEAM potential M1 is able to describe the elastic properties of the Li$_{x}$Si$_{y}$ phases properly and is thus suitable to be used for our proposed MD $C_{p}$ method.
Furthermore, we evaluated the radial pair distribution function from low temperature MD runs and compared our initial MD structures with the experimental ones (see supplementary material FIG. A1), where we achieved a good agreement. Thus our initial MD structures are valid for all further investigations. The final structure from the performed MD run is in good agreement with the averaged structure from each MD run, thus we can conclude that our MD runs are long enough (see supplementary material FIG. A1). 

\begin{table*}[t!!!!!!]
\setlength{\tabcolsep}{1.5pt}
\centering
\caption{Elastic properties of Li$_{x}$Si$_{y}$ phases. Here M1 and M2 are our MEAM calculations, DFT is our QUANTUM ESPRESSO LDA DFT calculation and Ref is the reference value (first row MEAM reference and second row ab-initio reference). The M2 setup was only used for the pure Si structure. Furthermore the subscripts a stands for the EVO Li$_{13}$Si$_{4}$, whereas b stands for the ICSD Li$_{13}$Si$_{4}$ structure. All values are given in GPa.\newline}
\tiny
\centering
\begin{tabular}{c|cc|cccc|c|ccccc|cccc|cc|cc|cccc|cc}
%\hline
  & \multicolumn{2}{c|}{Li} & \multicolumn{4}{c|}{Li$_{12}$Si$_{7}$} & Li$_{7}$Si$_{3}$ & \multicolumn{5}{c|}{Li$_{13}$Si$_{4}$} & \multicolumn{4}{c|}{Li$_{15}$Si$_{4}$} & \multicolumn{2}{c|}{Li$_{21}$Si$_{5}$} & \multicolumn{2}{c|}{Li$_{17}$Si$_{4}$} & \multicolumn{4}{c|}{Li$_{22}$Si$_{5}$} & \multicolumn{2}{c}{Si} \\
  & M1 & DFT &  M1 & DFT & \multicolumn{2}{c|}{Ref \cite{LiSi_MEAM}} & M1 & M1$_{\text{a}}$ & M1$_{\text{b}}$ & DFT$_{\text{a}}$ & \multicolumn{2}{c|}{Ref \cite{LiSi_MEAM}} & M1 & DFT & \multicolumn{2}{c|}{Ref \cite{LiSi_MEAM}} & M1 & DFT & M1 & DFT & M1 & DFT & \multicolumn{2}{c|}{Ref \cite{LiSi_MEAM}} & M2 & DFT \\\hline
$B_{\text{EOS}}$   & 11.8 & 15.0 & 60.6 & 39.8 & & & 30.4 & 43.3 & 42.2 & 35.8 & & & 37.8 & 33.5 & & & 36.0 & 34.4 & 40.4 & 35.2 & 35.8 & 34.3 & & & 94.9 & 94.3 \\
$B_{\text{Voigt}}$ & 12.8 & 15.3 & 55.8 & 40.5 & & & 47.7 & 43.3 & 40.7 & 36.7 & & & 37.9 & 33.9 & & & 36.4 & 34.9 & 36.6 & 35.5 & 35.8 & 35.1 & & & 95.3 & 96.2 \\
$C_{11}$           & 16.4 & 16.3 &  87.3 & 100.3 & 87.0 & 92.0 & 99.3 & 84.7 & 75.2 & 87.6 & 75.0 & 74.0 & 47.7 & 50.3 & 48.0 & 47.0 & 56.5 & 65.5 & 64.8 & 60.4 & 51.6 & 55.0 & 52.0 & 46.0 & 122.9 & 161.3 \\
$C_{22}$           & 16.4 & 16.3 &  79.5 & 106.5 & 78.0 & 97.0 & 99.3 & 72.7 & 69.9 & 80.4 & 70.0 & 61.0 & 47.7 & 50.3 & 48.0 & 47.0 & 56.5 & 65.5 & 66.2 & 60.4 & 51.6 & 55.0 & 52.0 & 46.0 & 122.9 & 161.3 \\
$C_{33}$           & 16.4 & 16.3 &  86.9 & 101.5 & 85.0 & 90.0 & 92.3 & 78.5 & 74.1 & 95.2 & 74.0 & 77.0 & 47.7 & 50.3 & 48.0 & 47.0 & 56.5 & 65.5 & 67.6 & 60.4 & 51.6 & 55.0 & 52.0 & 46.0 & 122.9 & 161.3 \\
$C_{12}$           & 11.0 & 14.8 &  44.0 & 5.7 & 45.0 & 5.0 & 36.9 & 24.7 & 21.7 & 18.6 & 22.0 & 17.0 & 32.9 & 25.7 & 33.0 & 21.0 & 26.3 & 19.7 & 23.3 & 23.1 & 27.9 & 25.1 & 28.0 & 23.0 & 81.4 & 63.6 \\
$C_{13}$           & 11.0 & 14.8 &  40.6 & 13.5 & 42.0 & 11.0 & 16.1 & 22.5 & 23.7 & 7.2 & 24.0 & 11.0 & 32.9 & 25.7 & 33.0 & 21.0 & 26.3 & 19.7 & 21.1 & 23.1 & 27.9 & 25.1 & 28.0 & 23.0 & 81.4 & 63.6 \\
$C_{23}$           & 11.0 & 14.8 &  39.7 & 8.7 & 39.0 & 8.0 & 16.1 & 29.7 & 28.0 & 7.6 & 28.0 & 10.0 & 32.9 & 25.7 & 33.0 & 21.0 & 26.3 & 19.7 & 20.9 & 23.1 & 27.9 & 25.1 & 28.0 & 23.0 & 81.4 & 63.6 \\
$C_{44}$           & 11.1 & 14.5 &  17.1 & 31.2 & 17.0 & 28.0 & 21.5 & 20.8 & 20.3 & 25.5 & 20.0 & 23.0 & 22.7 & 34.4 & 23.0 & 28.0 & 21.7 & 33.0 & 17.6 & 38.0 & 29.6 & 41.7 & 30.0 & 35.0 & 81.3 & 76.7 \\
$C_{55}$           & 11.1 & 14.5 &  18.8 & 32.0 & 20.0 & 26.0 & 21.5 & 18.1 & 13.6 & 28.0 & 14.0 & 24.0 & 22.7 & 34.4 & 23.0 & 28.0 & 21.7 & 33.0 & 17.5 & 38.0 & 29.6 & 41.7 & 30.0 & 35.0 & 81.3 & 76.7 \\
$C_{66}$           & 11.1 & 14.5 &  12.4 & 27.1 & 15.0 & 24.0 & 31.2 & 16.7 & 13.6 & 32.1 & 14.0 & 28.0 & 22.7 & 34.4 & 23.0 & 28.0 & 21.7 & 33.0 & 18.9 & 38.0 & 29.6 & 41.7 & 30.0 & 35.0 & 81.3 & 76.7 \\%\hline
\end{tabular}
\label{elastic_properties}
\end{table*}

%$B_{\text{cubic}}$ & [GPa] & 12.8 & 15.3 & 58.4 & 37.2 & & & 57.7 & 44.7 & 39.5 & 41.6 & & & 37.9 & 33.9 & & & 36.4 & 34.9 & 37.1 & 35.5 & 35.8 & 35.1 & & & 95.3 & 96.2 \\

\subsection{Results II - specific heat capacities\label{results2}} 

\subsubsection{Specific heat of lithium and silicon}

The experimental reference data for the isobaric heat capacity was obtained by \citet{LiSi_exp}. 
Initial calculations were carried out on the pure Li and Si systems to ensure that the MD results  
agree with the ones obtained by DFT and with the experimental data.
This comparison is the first evaluation of the MEAM potentials for the calculation of thermodynamic properties
within our MD approach.

\begin{figure}[h!]
\centering
 \includegraphics[width=0.47\textwidth]{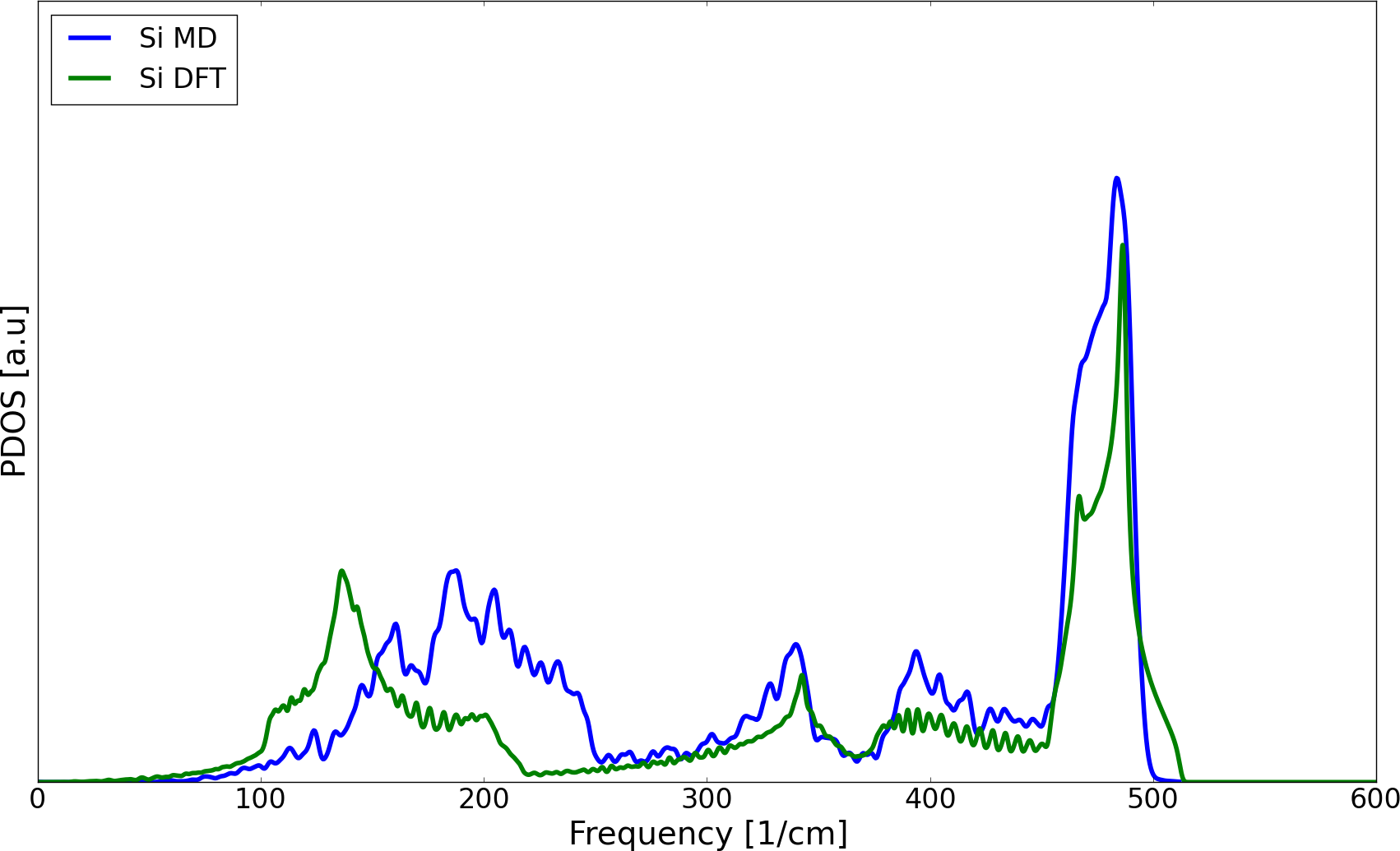}
 \caption{Phonon density of states of silicon.}
 \label{PDOS_Si}
\end{figure}

One core property for a good description of the heat capacities is the phonon density of states. 
We calculated the PDOS for pure silicon with DFT using QUANTUM ESPRESSO and from MD using LAMMPS with the "fix\_phonon" option.
The results of these calculations are visualized in FIG. \ref{PDOS_Si}. The resulting DFT and MD PDOS are in good agreement, because 
the main features appear in both density of states. Based on this result we continued to use DFT (in the framework of the quasi-harmonic approximation)  
and our MD $C_{p}$ method to calculate the isobaric heat capacity $C_{p}$ for silicon and lithium. 

\begin{figure}[h!]
  \subfigure[~$g(r)$]{\includegraphics[width=0.47\textwidth]{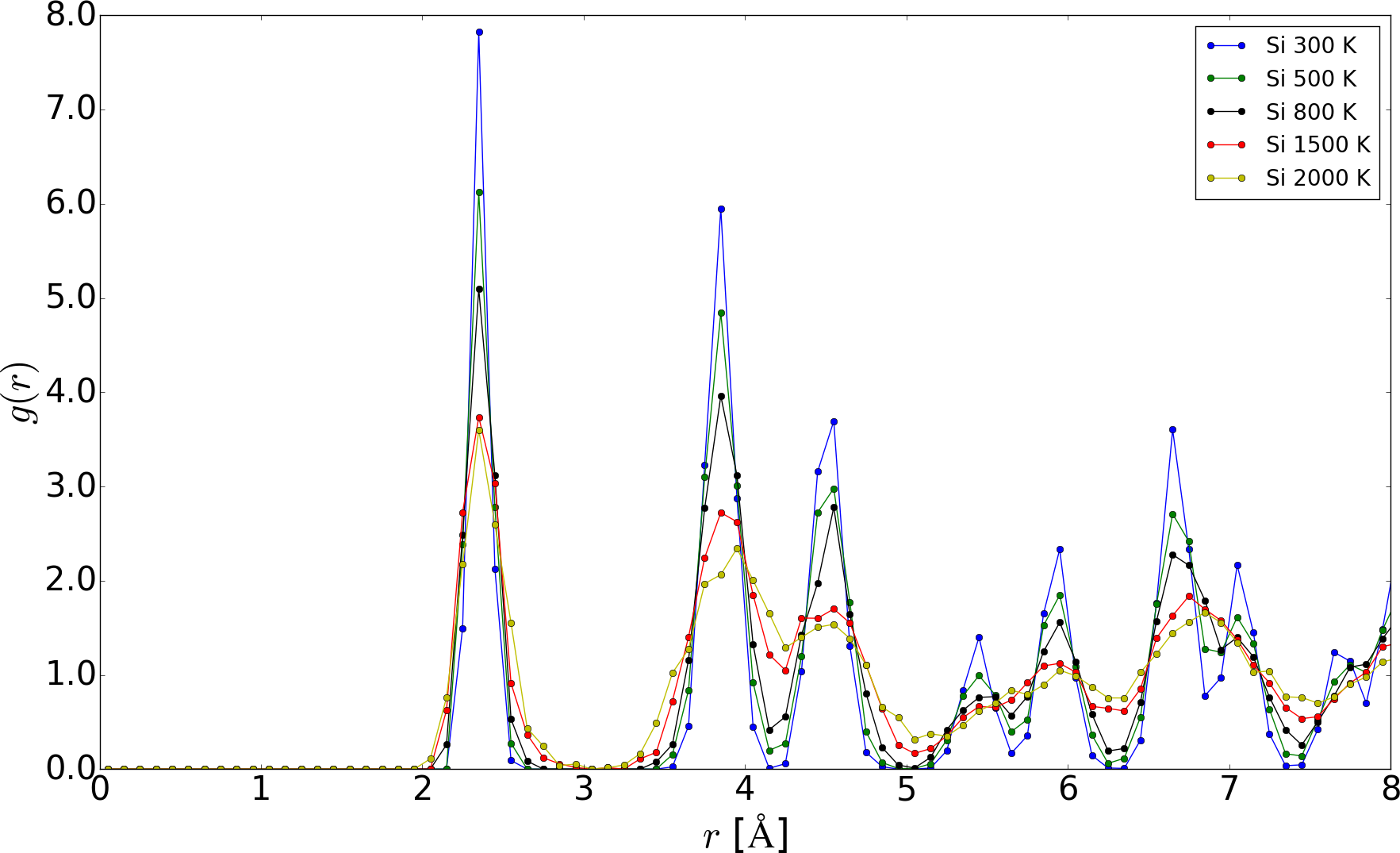}}
 \hspace{1em}
 \subfigure[~$C_{p}$]{\includegraphics[width=0.47\textwidth]{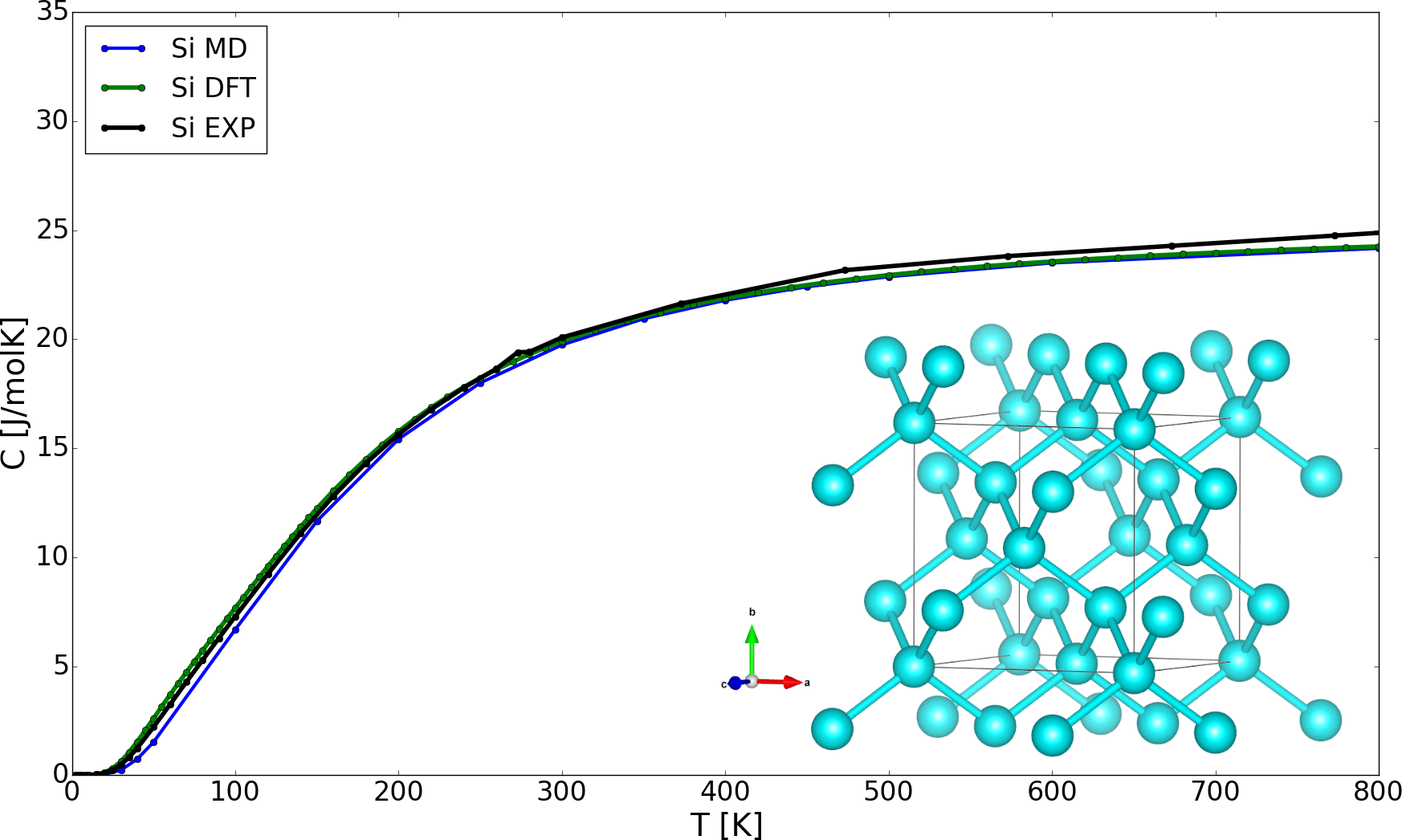}}
 \caption{(a) Radial pair distribution function $g(r)$ and (b) specific heat capacity  $C_{p}$ of silicon. Inset shows crystalline structure with Si-Si bonds visualizing Si-structure elements (Si-tetrahedron). The experimental $C_{p}$ data is extracted from \citet{Si_cp_exp_T_down_300K} for $T < 300$ K and 
 from \citet{Si_cp_exp_T_over_300K} for $T \geq 300$ K.}
 \label{Cp_Si}
\end{figure}

\begin{figure}[h!]
  \subfigure[~$g(r)$]{\includegraphics[width=0.47\textwidth]{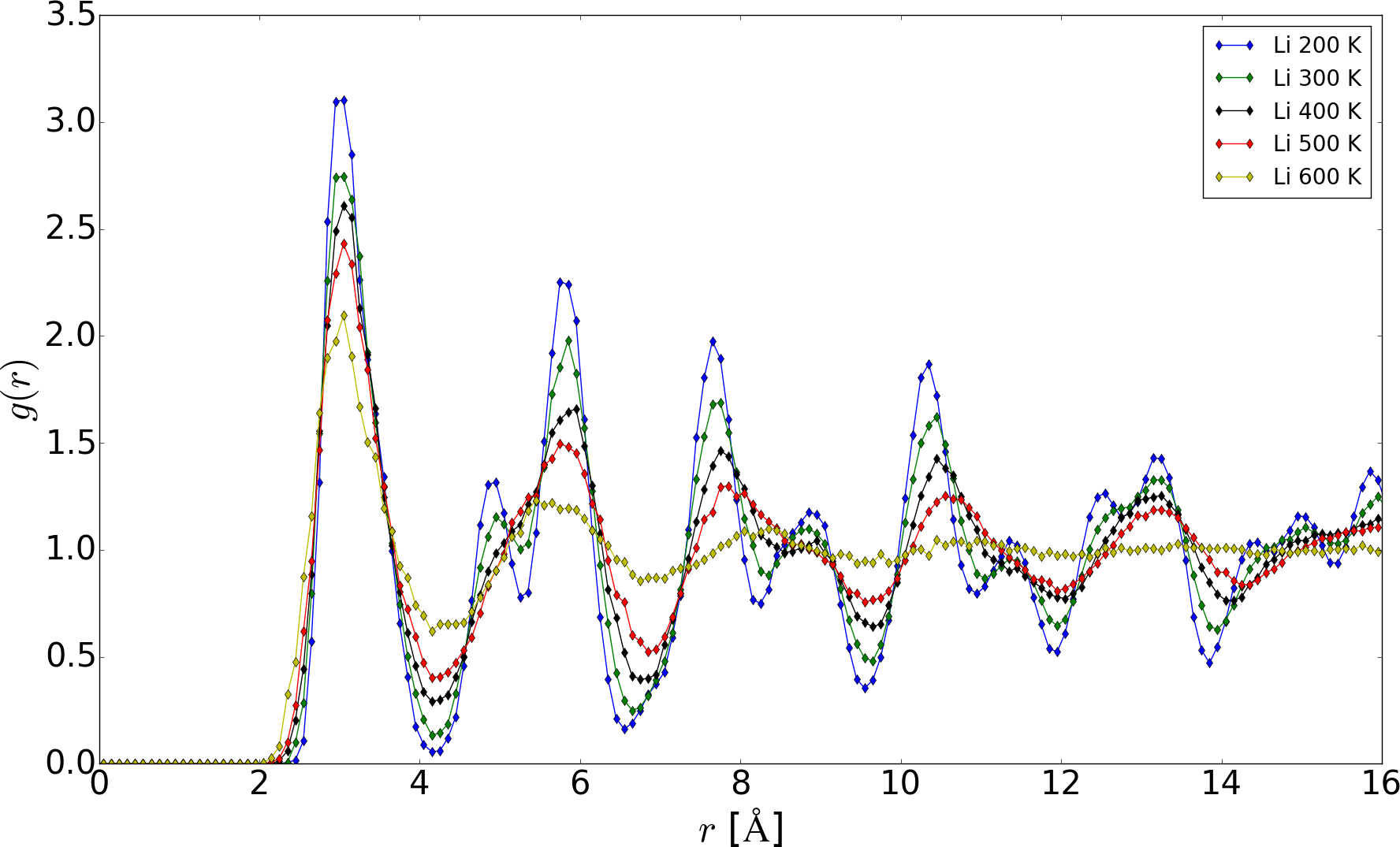}}
 \hspace{1em}
 \subfigure[~$C_{p}$]{\includegraphics[width=0.47\textwidth]{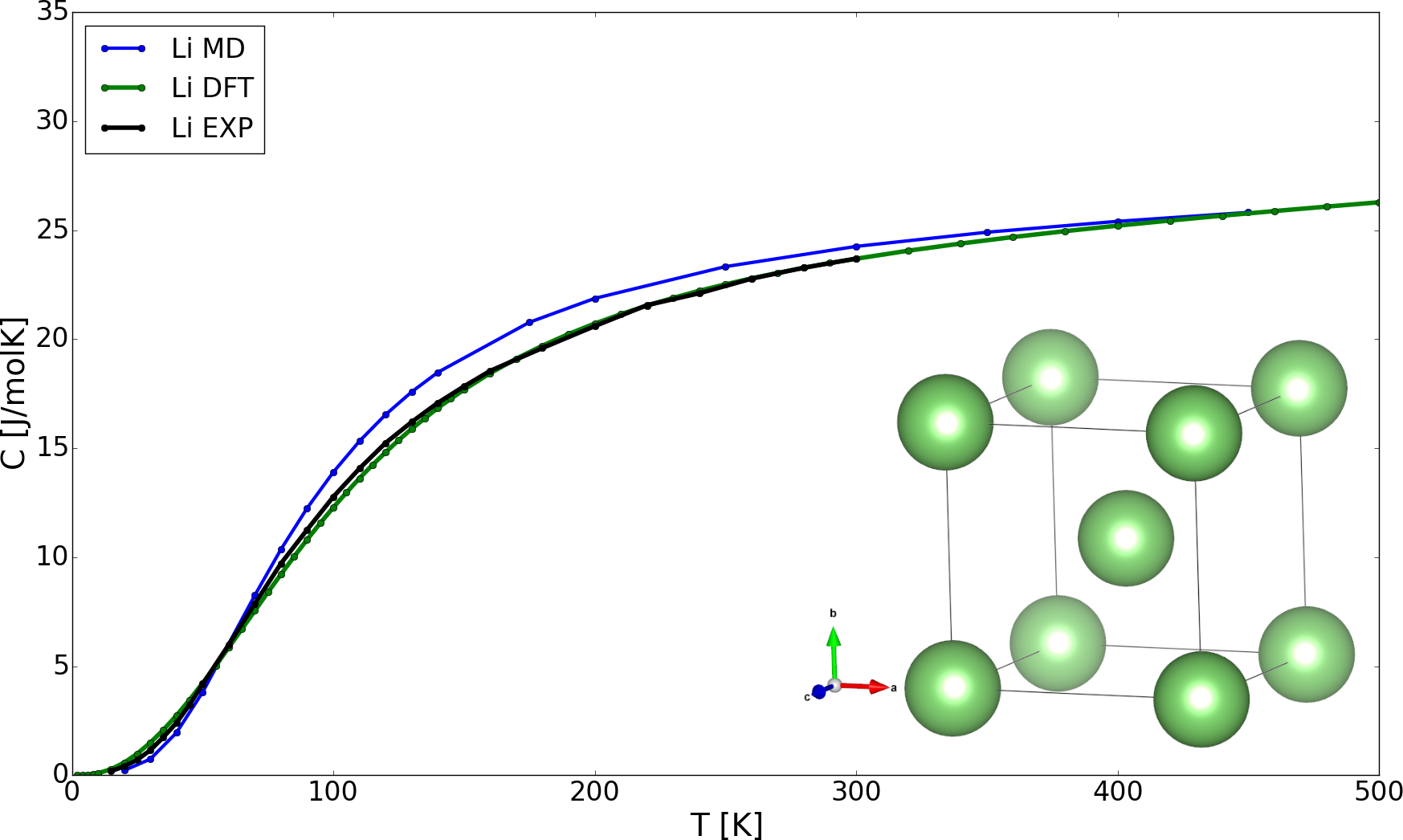}}
 \caption{(a) Radial pair distribution function $g(r)$ and (b) specific heat capacity $C_{p}$ of lithium. Inset shows crystalline structure. The experimental $C_{p}$ data is extracted from \citet{Li_cp_exp_Simon_Swain_data} (original published by \citet{Li_cp_exp_Simon_Swain_origin}).}
 \label{Cp_Li}
\end{figure}

In FIG. \ref{Cp_Si}~(b) and \ref{Cp_Li}~(b) the isobaric heat capacities for pure silicon and lithium are plotted versus temperature.
In the case of silicon (see FIG. \ref{Cp_Si}~(b)), the values of our MD calculations agree with the DFT and experimental values over a wide range of temperatures. 
FIG. \ref{Cp_Si}~(a) shows the RPDF of pure Si. The crystallinity of this system can be clearly seen, even at rather high temperatures. This indicates that the Si itself is thermally very stable. However, from 1500 K on the 
features at larger distances smear out, indicating a decrease in crystallinity at such temperatures.
It should be noted here that the isobaric specific heat capacity for lithium differs in the temperature region 
from 100 to 200 K slightly from the DFT values, which could lead to small systematic errors for the 
Li$_{x}$Si$_{y}$ phases in those temperature regions due to the given MEAM potential. 
However, there is a general agreement of our MD results with those obtained by DFT and the experiment for pure lithium in the entire temperature range.
As can be seen in FIG. \ref{Cp_Li}~(a), the crystallinity of lithium is preserved up to a temperature of ca. 500 K. Clearly at higher temperatures ($> 600$ K) the RPDF shows liquid behavior and the crystalline features become absent, indicating a melting process appearing at around 500 K.
Based on these preliminary results our MD approach should allow to calculate the specific heat capacity for the Li$_{x}$Si$_{y}$ phases. 
\subsubsection{Specific heat capacity of lithium silicides} 
\begin{figure}[t!]
  \subfigure[~$g(r)$]{\includegraphics[width=0.47\textwidth]{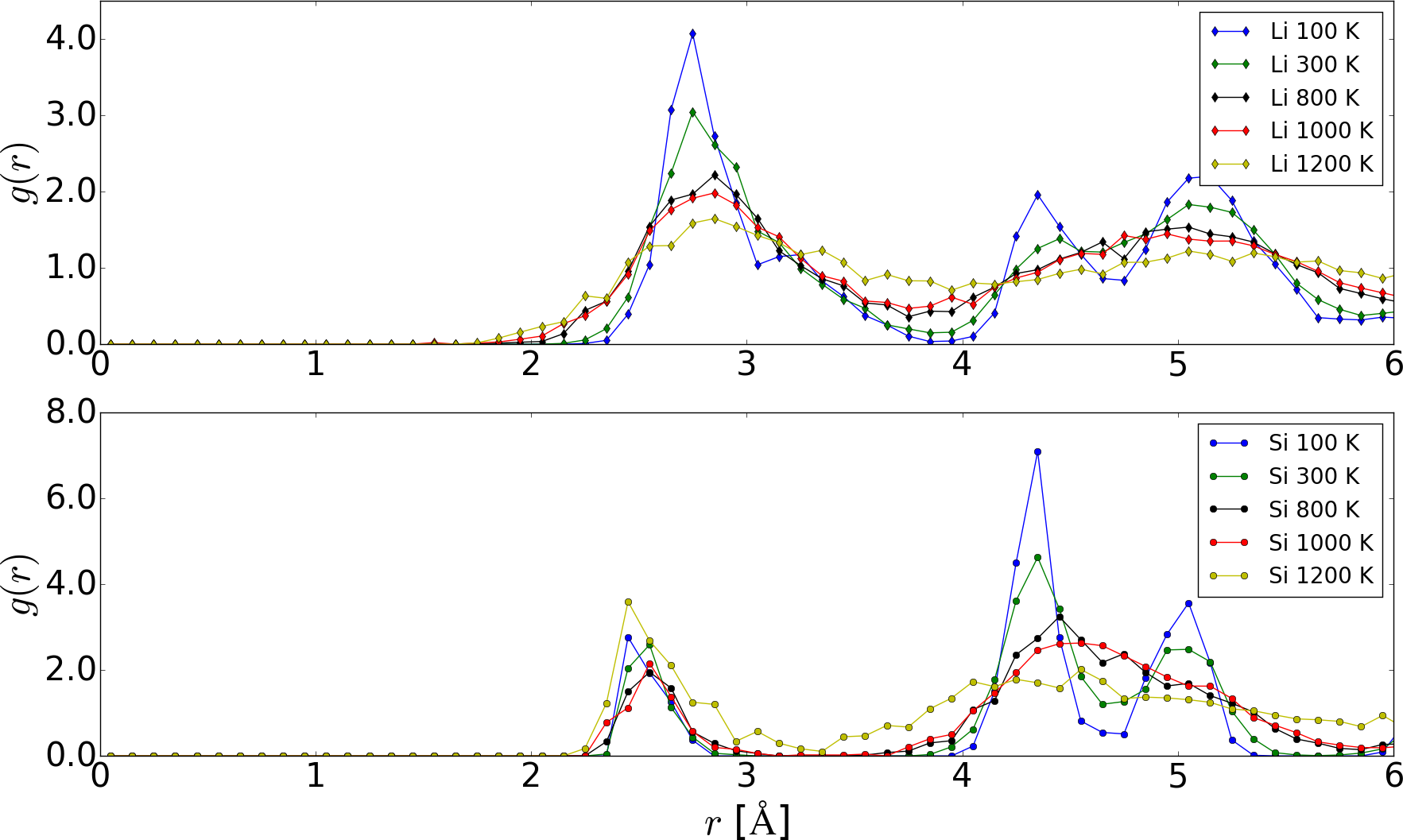}}
 \hspace{1em}
 \subfigure[~$C_{p}$]{\includegraphics[width=0.47\textwidth]{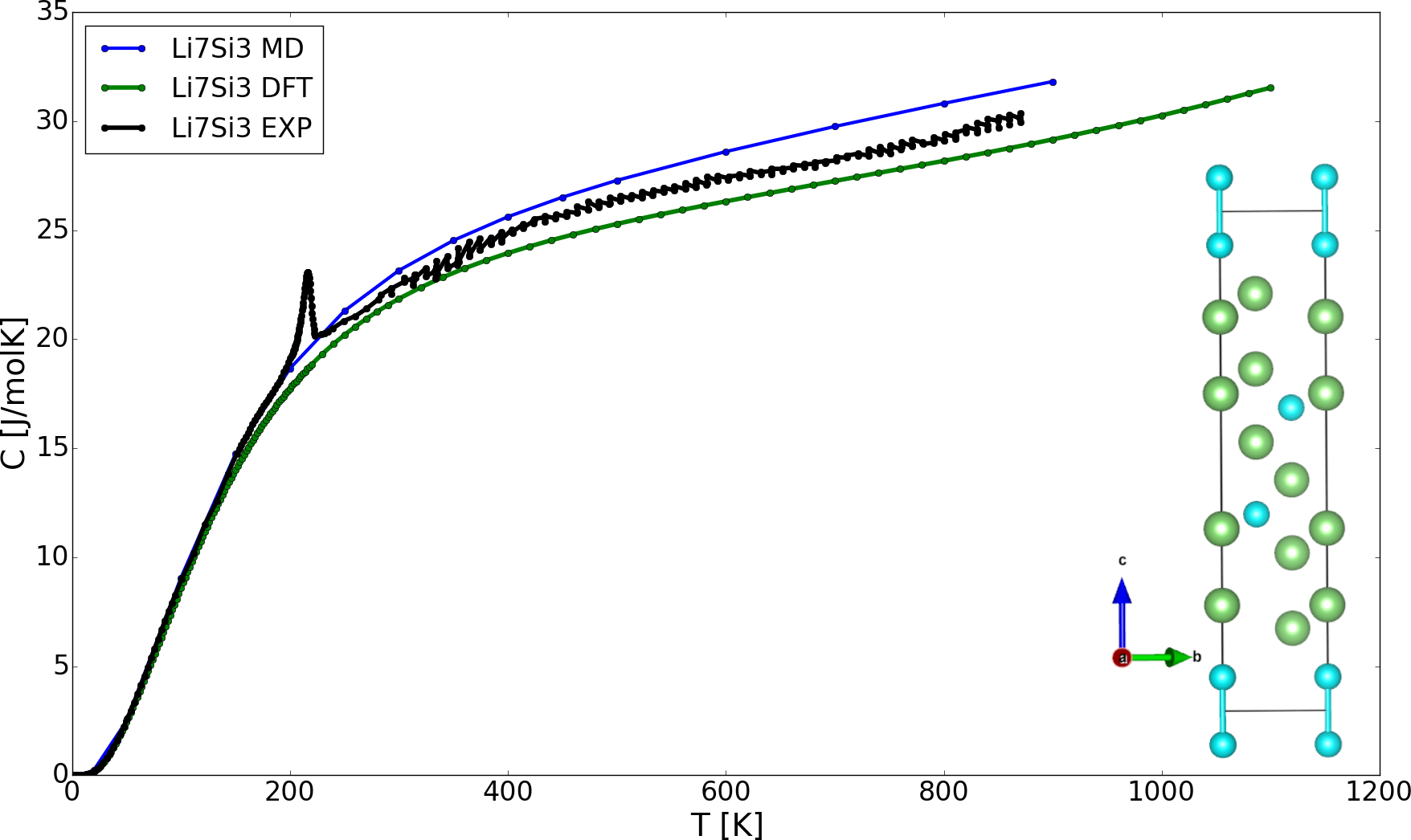}}
 \caption{(a) Radial pair distribution function $g(r)$ and (b) specific heat capacity $C_{p}$ of Li$_{7}$Si$_{3}$. Inset shows crystalline structure with Si-Si bonds visualizing Si-structure elements (Si-dimers). The experimental $C_{p}$ data is extracted from \citet{LiSi_exp}.}
 \label{Cp_Li7Si3}
\end{figure}

In general all our calculated ab-initio isobaric heat capacity curves for the Li$_{x}$Si$_{y}$ phases are in good agreement with experimental curves
over the entire temperature range. Only for Li$_{12}$Si$_{7}$ and Li$_{13}$Si$_{4}$ the DFT values differ from the experimental values in the 
high temperature region, which indicates that for these phases high-temperature effects may play a role. As already mentioned earlier these 
effects cannot be described by the quasi-harmonic approximation and therefore cannot be treated within this framework. For these phases 
our proposed MD $C_{p}$ method seems to give a better description of the high-temperature limit.   
For Li$_{7}$Si$_{3}$, as seen in FIG. \ref{Cp_Li7Si3}~(b), the heat capacity calculated with the MEAM potentials is in good agreement with experimental and DFT results. 
There is a slight overestimation at higher temperatures, but the trend is preserved for both methods. A possible explanation could be that silicon forms stable dimers \cite{LiSiZintl3} in Li$_{7}$Si$_{3}$, which show local order even at higher temperatures. \\
In FIG. \ref{Cp_Li7Si3}~(a) the RPDF of Li and Si within Li$_{7}$Si$_{3}$ is displayed. For lithium a very similar trend as in pure lithium is observed, that is the 'double peak' at approximately 4-5 \AA, which smears out at higher temperatures. The general features for low distances are preserved as well. At $T >$ 1000 K the RPDF for Li shows liquid features, while the crystalline ones are reduced. This fact indicates a melting process at this temperature. For Si, the RPDF is fairly constant over the investigated temperature range, indicating the high stability of the Si structure within the phase.
% double peak at Si -> why?
\begin{figure}[t!]
  \subfigure[~$g(r)$]{\includegraphics[width=0.47\textwidth]{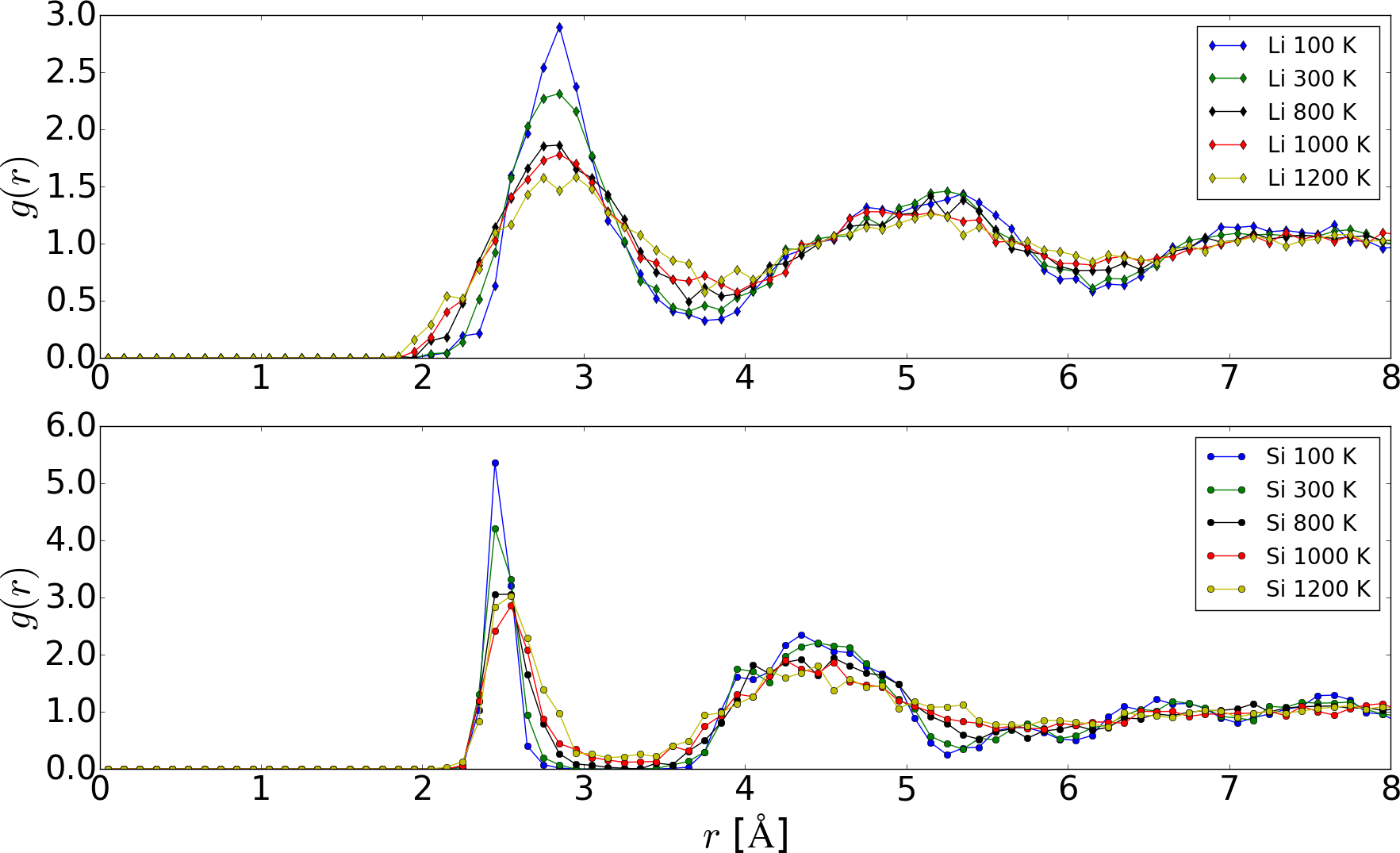}}
 \hspace{1em}
 \subfigure[~$C_{p}$]{\includegraphics[width=0.47\textwidth]{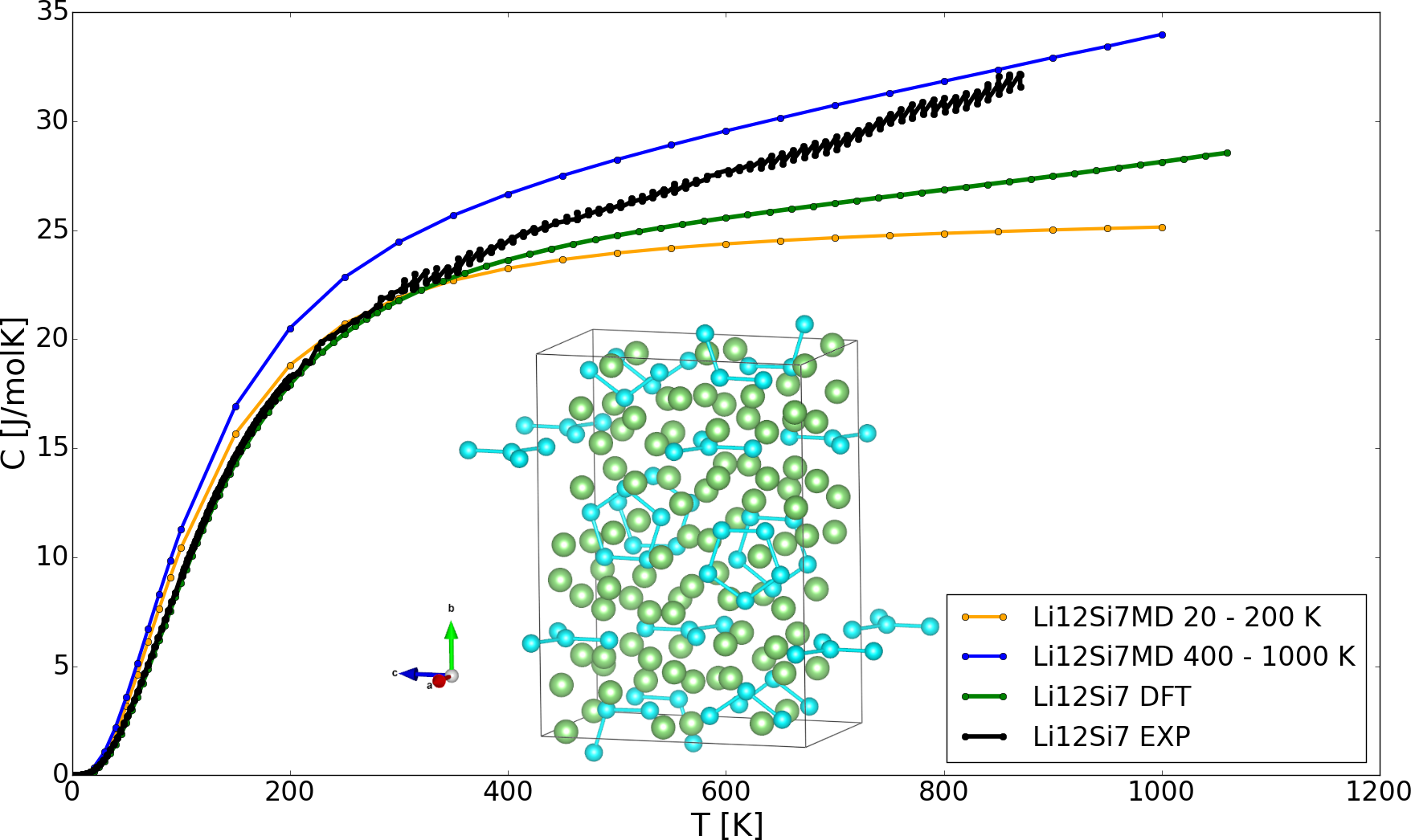}}
 \caption{(a) Radial pair distribution function $g(r)$ and (b) specific heat capacity $C_{p}$ of Li$_{12}$Si$_{7}$. Inset shows crystalline structure with Si-Si bonds visualizing Si-structure elements (Si-pentagon-ring, Si-star). The experimental $C_{p}$ data is extracted from \citet{LiSi_exp}.}
 \label{Cp_Li12Si7}
\end{figure}
\noindent
\newline 
For Li$_{12}$Si$_{7}$ in FIG. \ref{Cp_Li12Si7}~(b) we observed two different $V(T)$ dependencies, one occurring at lower temperatures (up to approximately 400 K) leading to a specific $\alpha_{\text{low}-T}$ and one occurring at higher temperatures, which results in a different $\alpha_{\text{high}-T}$ (see TAB. \ref{table_alpha}).
Based on the Li RPDF (see FIG. \ref{Cp_Li12Si7}~(a)), which indicates structural changes at temperatures $T > 300$ K we separated two temperature regions on from 20 - 200 K leading to $\alpha_{\text{low}-T}$ and another from 400 - 1000 K leading to $\alpha_{\text{high}-T}$.
This difference may be caused by anharmonic contributions or diffusion effects which take place at higher temperatures, giving rise to a different expansion behavior. The two distinguishable values of $\alpha$ lead to different $C_{p}$ values, see Eq. (\ref{equation_cp}). It is not possible to describe phase transitions with the standard quasi-harmonic approximation, which delivers a possible explanation why the DFT data are in good agreement with experiments in the low temperature region but not in the high temperature region. In the Li$_{12}$Si$_{7}$ phase the silicon atoms form Si$_5$ rings with a lithium atom inside. These Si$_{5}$ rings are stacked between two Li$_{5}$ rings. Concerning this situation, it is not necessary to break a bond to disturb the structure, as the Li$_{5}$ rings can slide between the Si$_{5}$ and the crystalline structure may transform into another one or becomes amorphous. \\
FIG. \ref{Cp_Li12Si7}~(a) displays the RPDF of Li and Si within Li$_{12}$Si$_{7}$. For lithium a somewhat different behavior in comparison to pure Li is found, as there is only one peak at a distance of 4-5 \AA. Starting from ca. 1000 K, the features in the RPDF become more liquid-like. The shape and position of Si peaks are largely temperature independent up to high temperatures, showing again high stability of the Si structures.

\begin{figure}[t!]
  \subfigure[~$g(r)$]{\includegraphics[width=0.47\textwidth]{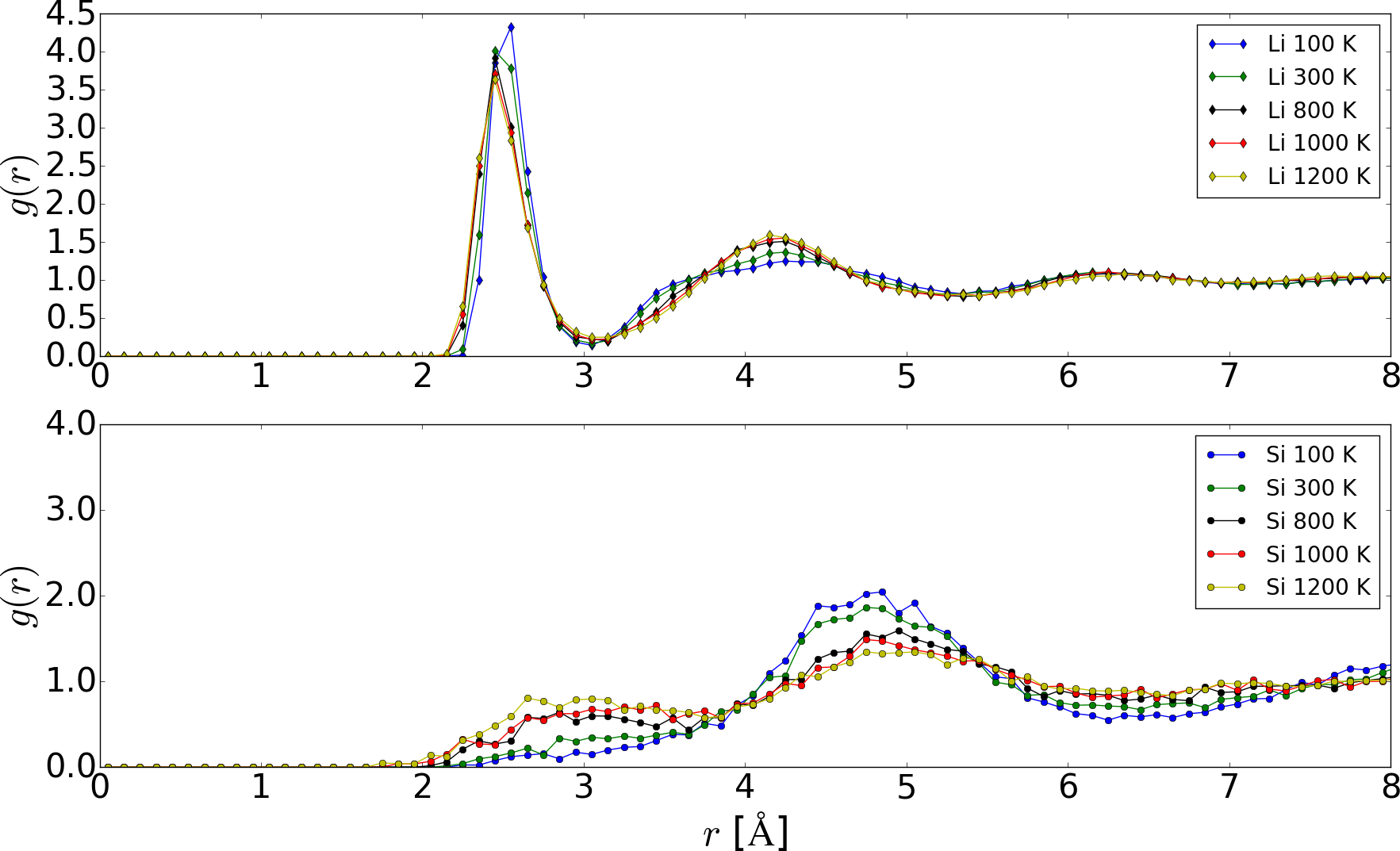}}
 \hspace{1em}
 \subfigure[~$C_{p}$]{\includegraphics[width=0.47\textwidth]{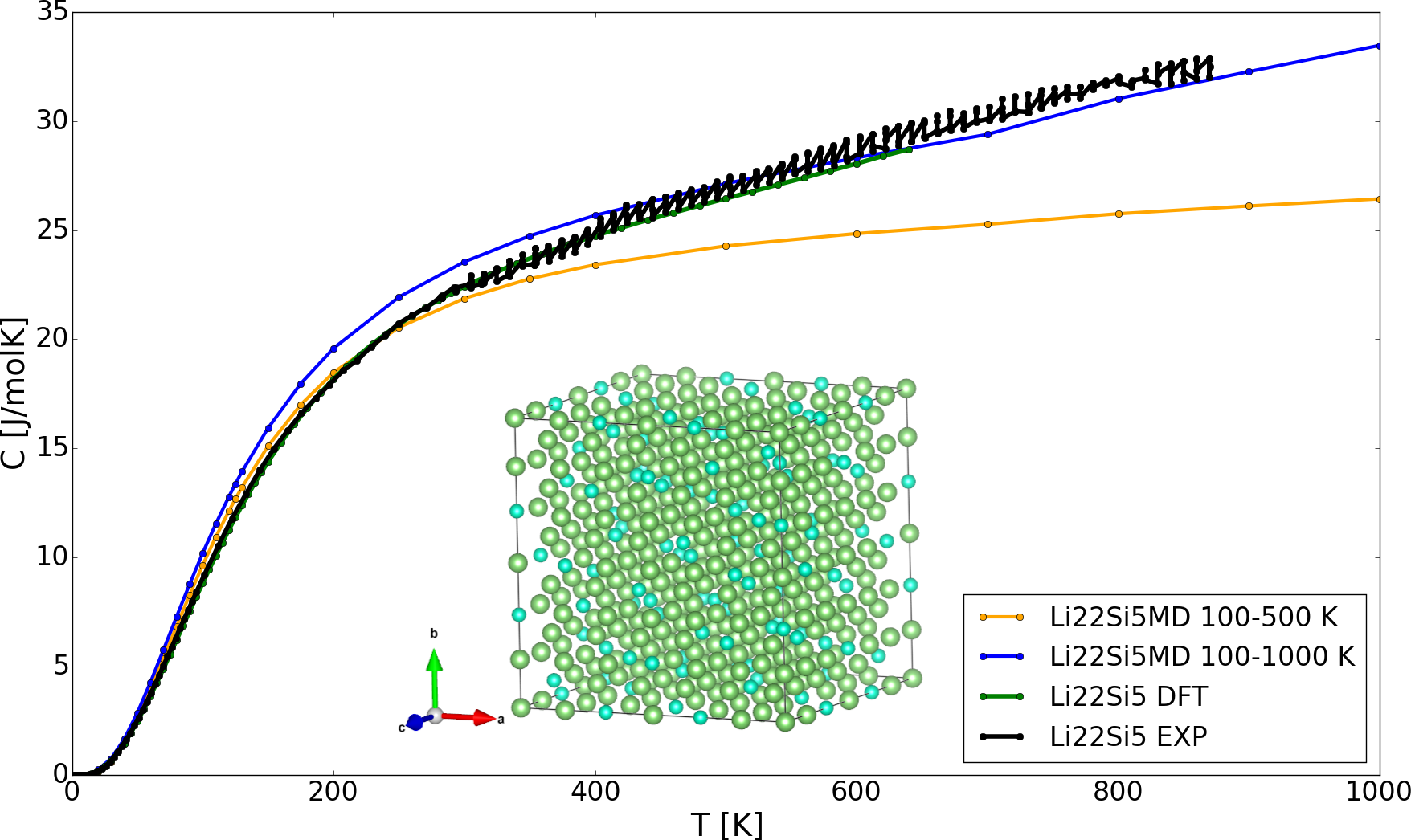}}
 \caption{(a) Radial pair distribution function $g(r)$ and (b) specific heat capacity $C_{p}$ of Li$_{22}$Si$_{5}$. Inset shows crystalline structure with Si-structure elements (Si-atoms). The experimental $C_{p}$ data is extracted from \citet{LiSi_exp}.}
 \label{Cp_Li22Si5}
\end{figure}

\begin{figure}[t!]
  \subfigure[~$g(r)$]{\includegraphics[width=0.47\textwidth]{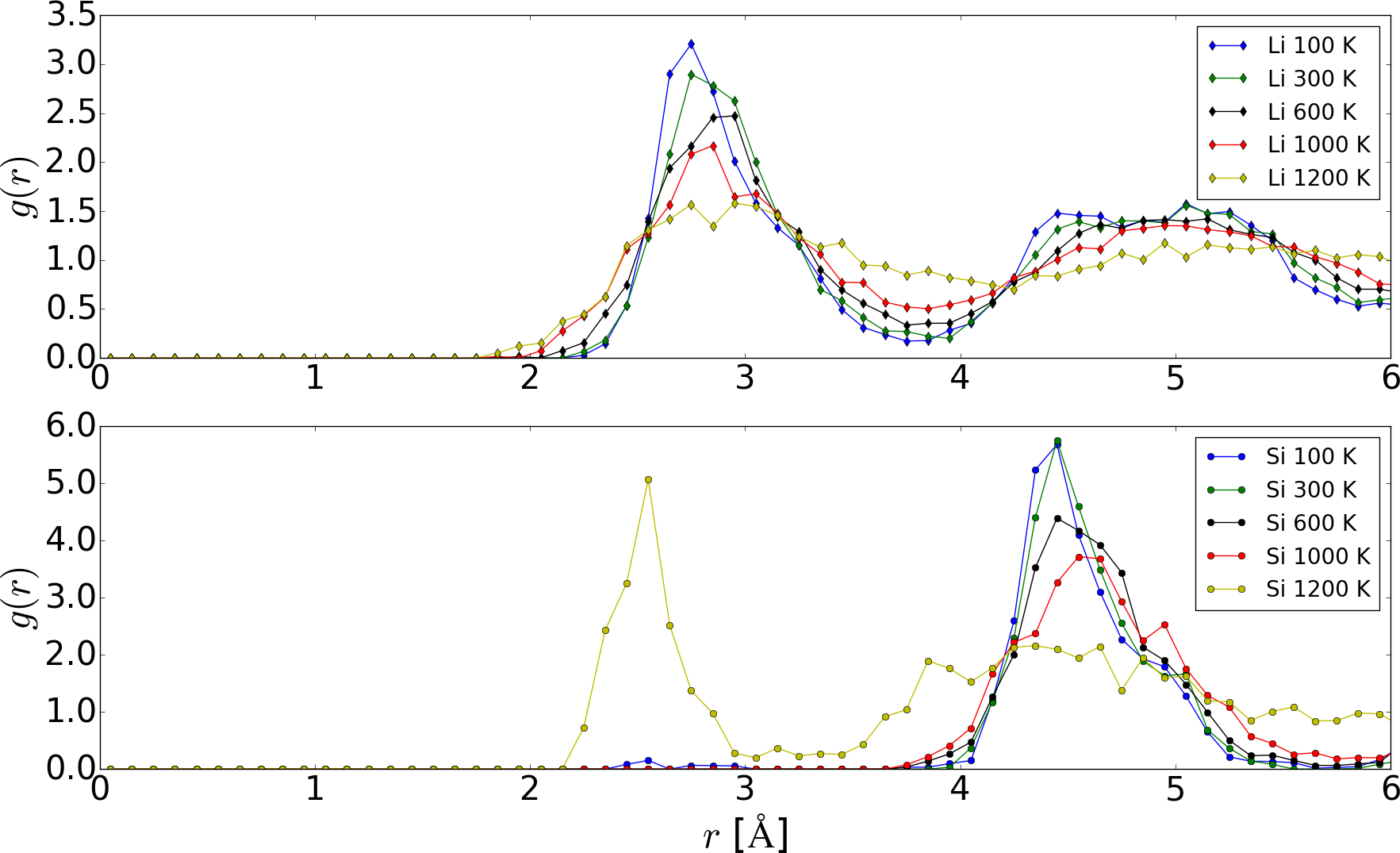}}
 \hspace{1em}
 \subfigure[~$C_{p}$]{\includegraphics[width=0.47\textwidth]{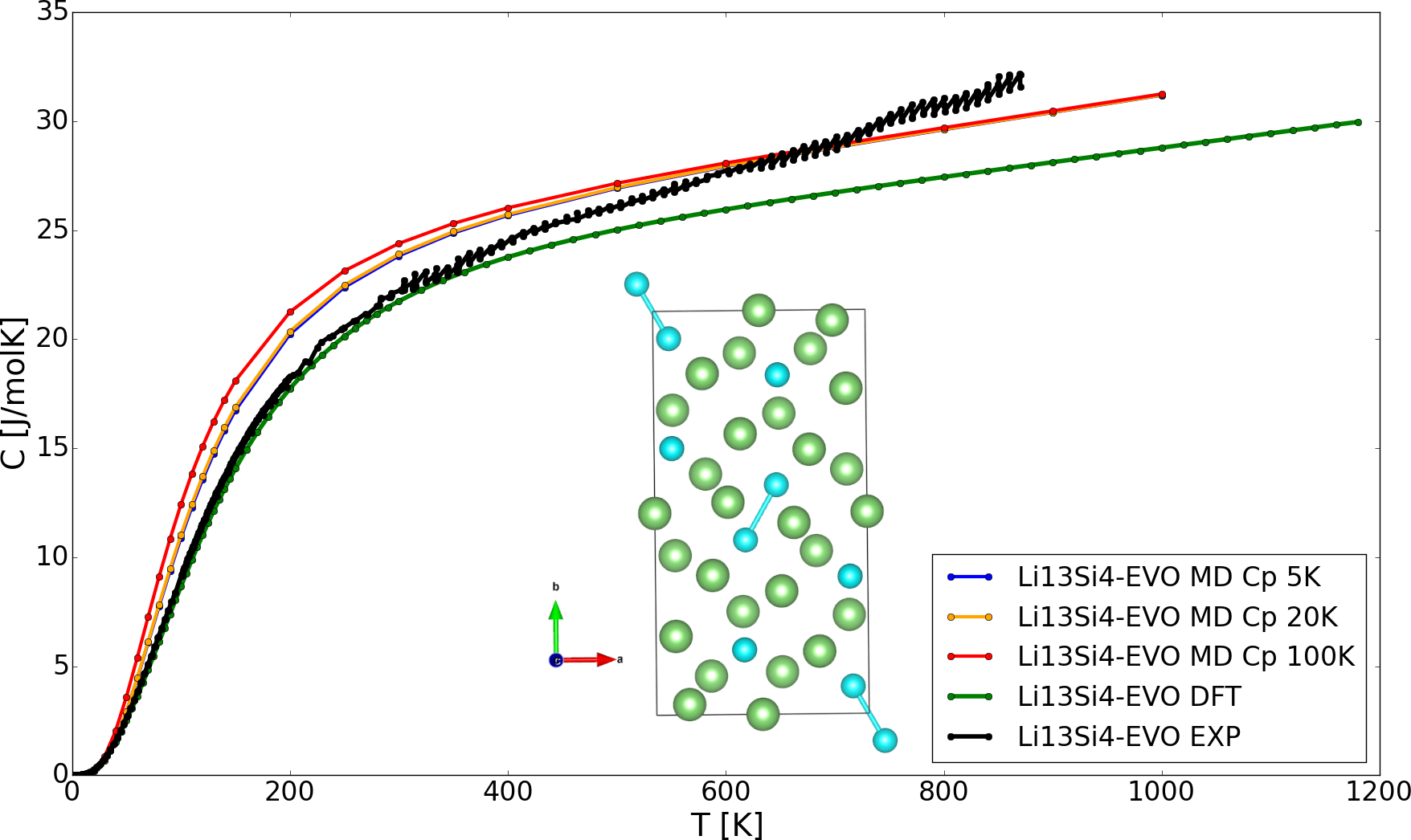}}
 \caption{(a) Radial pair distribution function $g(r)$ and (b) specific heat capacity $C_{p}$ of Li$_{13}$Si$_{4}$. Inset shows crystalline structure with Si-Si bonds visualizing Si-structure elements (Si-atoms, Si-dimers). The MD calculations have been performed for different low temperatures. The DFT values were already published in \cite{Li13Si4}. The experimental $C_{p}$ data is extracted from \citet{LiSi_exp}.}
 \label{Cp_Li13Si4}
\end{figure}

For Li$_{22}$Si$_{5}$ in FIG. \ref{Cp_Li22Si5}~(b) a similar behavior like for Li$_{12}$Si$_{7}$ has been found. 
Again, there are two different $\alpha$ values describing the resulting heat capacity at different temperature regimes.
One major difference is that the Li RPDF (see FIG. \ref{Cp_Li22Si5}~(a)) indicates no structural changes at higher temperatures, thus both linear $\alpha$ values were evaluated with the same starting temperature $T = 100$ K until 500 K for $\alpha_{\text{low}-T}$ and until 1000 K for $\alpha_{\text{high}-T}$.
Here, the agreement of the two different plots to the other values is more obvious. \\
FIG. \ref{Cp_Li22Si5}~(a) displays the RPDF of Li and Si within Li$_{22}$Si$_{5}$. 
The Li RPDF is somewhat different from those of the other phases. 
In detail, the initial sharp peak around 2.5 \AA \ looks crystalline, while no other crystalline features are present for the Li$_{22}$Si$_{5}$ structure. 
Thus the Li atoms prefer a specific Li-Li distance, which is constant over the investigated temperature range.
In addition, the following features have characteristics of a liquid (declining sinusoidal-like behavior). This indicates that even at low temperatures a permanent Li diffusion in the phase takes place. 
Experimental crystal structure determination methods like XRD or neutron scattering are therefore only able to measure snapshots of the structure at a specific time, at which the diffusive Li atoms are quasi-frozen for the moment of the measurement.  
In addition, the liquid-like behavior will make it very difficult to distinguish between Li$_{22}$Si$_{5}$, Li$_{21}$Si$_{5}$, and Li$_{17}$Si$_{4}$ as all of them have a high and very similar Li content.
For Si on the other hand there is a very broad peak at around 5 \AA \ at low temperatures, indicating that the Si atoms within this phase are only ordered to a certain degree. Once the temperature is increased, the peak strongly smears out, thus the Si atoms become disordered. This is obvious, as within this phase Si is not present in clusters, but only as single atoms. 
These atoms are surrounded by a large amount of diffusive Li. With that, the Si atoms cannot form any ordered structures and stay as single atoms.

For Li$_{13}$Si$_{4}$ in FIG. \ref{Cp_Li13Si4}~(b) a temperature dependent PDOS was calculated (5 K, 20 K, 100 K). 
Here the agreement with experiment and DFT $C_{p}$ results for a low temperature PDOS is better. 
This might be caused by suppressed diffusion processes \cite{Li13Si4} within the phase at low temperature, which are certainly more important for higher temperatures. 
Thus the description of the systems becomes better at low temperatures because we "freeze out" the diffusion process. 
In FIG. \ref{Cp_Li13Si4}~(a) the RPDF of Li and Si within Li$_{13}$Si$_{4}$ is shown. The first peak for Li shows similarities to Li$_{12}$Si$_{7}$ and Li$_{7}$Si$_{3}$. However, in contrast to 
Li$_{7}$Si$_{3}$ and in correspondance to Li$_{12}$Si$_{7}$, there is a single, broad second peak at around 4.5 \AA. The RPDF for Si 
indicates that initially more single atoms than single dimers are present, which can be seen from the broad peak at approximately 4.5 \AA \ and from the crystal structure itself. However, at a temperature of 1200 K, this peak is strongly smeared out and an additional peak at about 2.5 \AA \ appears, again resembling the features seen in all other phases. At this temperature the first nearest neighbors peak in the Si RPDF increases significantly, which indicates structural changes concerning the Si atoms with increasing temperature.

\begin{table}[ht!]
\setlength{\tabcolsep}{1.5pt}
\tiny
\centering
\caption{Thermal expansion coefficients for Li$_{x}$Si$_{y}$ crystal structures. If more than one value is present, the temperature regime is given.\newline} 
\centering
\begin{tabular}{l|ccc}
\multirow{2}{*}{Li$_x$Si$_y$} & $\alpha_{l,\text{low}-T}$ & $\alpha_{l,\text{high}-T}$ & $\alpha_{l,\text{ref}}$ \cite{thermal_expansion} \\
	   & [$10^{-6}$ K$^{-1}$] & [$10^{-6}$ K$^{-1}$] & [$10^{-6}$ K$^{-1}$] \ \\ \hline

%\multirow{3}{*}{Li$_x$Si$_y$} & \multicolumn{2}{c}{$\alpha_{l,\text{NPT-MD}}$} & \multirow{2}{*}{$\alpha_{l,\text{ref}}$}\cite{thermal_expansion}  \ \\
% & $\alpha_{l,\text{low}-T}$ & $\alpha_{l,\text{high}-T}$ & \\
%	   & [$10^{-6}$ K$^{-1}$] & [$10^{-6}$ K$^{-1}$] & [$10^{-6}$ K$^{-1}$] \ \\ \hline
Si & 5.6  & 5.6 & 2.6–3 \ \\
\multirow{2}{*}{Li$_{12}$Si$_{7}$} & 4.4 & 18.7 & \multirow{2}{*}{---} \ \\
& (20-200 K) &  (400-1000 K) & \ \\
Li$_{7}$Si$_{3}$ & 26.5 & 26.5 & --- \ \\ 
Li$_{13}$Si$_{4}$& 58.6 & 58.6 & --- \ \\
\multirow{2}{*}{Li$_{22}$Si$_{5}$}& 33.9 & 72.6 & \multirow{2}{*}{---} \ \\
& (100-500 K) & (100-1000 K) & \ \\
Li & 54.7 & 54.7 & 56 \ \\
\end{tabular}
\label{table_alpha}
\end{table}

% General summary 
%\newpage
As one can see, the isobaric heat capacity in low temperature region (0 - 400 K) calculated with DFT in the framework of the quasi-harmonic approximation 
is in excellent agreement with experimental data. In the high temperature region (400 - 1200 K) anharmonic effects become important and the 
quasi-harmonic approximation is not able to produce the right thermal expansion. 
Using our proposed MD $C_{p}$ method calculation of the isobaric heat capacity we are able to
deliver improved results in the high temperature region (400 - 1200 K) in respect to the reference experiment. There are slight deviations in the low temperature region compared to the experiment or the DFT results. The use of both theoretical methods enables us to give a good description of the isobaric heat capacity for the Li$_{x}$Si$_{y}$ crystal structures over the entire temperature region. 

\section{Computational time analysis\label{computational_time_analysis}}

DFT in the framework of the quasi-harmonic approximation is able to determine quite accurate and reasonable isochoric heat capacity values, 
but at which cost? This computational framework is expensive, in respect to computational time (see TAB. \ref{table_qe_time}). 
Systems with a relative small number of atoms in their unit cell (e.g. Li, Si) can be treated within this framework.
In contrast, a large number of atoms (e.g. lithium silicides) leads to a dramatically increased computational time (see TAB. \ref{table_qe_time}). 
For example, the computational time to calculate the isochoric heat capacity for the lithium rich Li$_{22}$Si$_{5}$ phase is over 500000 CPUh. 
Our proposed MD $C_{p}$ method is based on force-fields. For the same Li$_{22}$Si$_{5}$ phase,
we only need about 61 CPUh to get access to reasonable values for the isobaric heat capacity. 
Given these numbers our method is more than 4 orders of magnitude faster than ab-initio computational thermodynamics 
(see TAB. \ref{table_lammps_time}). In summary, our MD $C_{p}$ method based computational thermodynamics allows to describe complex structures with many atoms per unit cell in reasonable computational time. 

\begin{table}[h!]
\setlength{\tabcolsep}{1.5pt}
\tiny
\centering
\caption{CPU time using DFT for Li$_{x}$Si$_{y}$ crystal structures. The time of self-consistent-field calculation is $t_{\text{SCF}}$, the number of cores used for the SCF calculation is $N_{\text{cores}}^{\text{SCF}}$,
the time for one displacement is $t_{\text{disp}}$, the number of cores used for the displacement calculations is $N_{\text{cores}}^{\text{disp}}$,
the number of calculated volumes is $N_{\text{vol}}$, the number of displacements is $N_{\text{disp}}$ and the time for one quasi-harmonic-approximation (QHA) run per structure is $t_{\text{QHA}}$.\newline}
\centering
\begin{tabular}{l|ccccccc}
System & $t_{\text{SCF}}$ & $N_{\text{cores}}^{\text{SCF}}$ & $t_{\text{disp}}$ & $N_{\text{cores}}^{\text{disp}}$ & $N_{\text{vol}}$& $N_{\text{disp}}$ & $t_{\text{QHA}}$  \ \\ 
& [CPUh] &  &[CPUh] & & & &[CPUh]\ \\ \hline
Li      &0.1 &   3  &  3   &   3 & 5 &   2 &    91.5  \ \\
%LiSi    &  2 &   3  &  6   &   6 & 5 &  12 &  2190   \ \\
Li$_{12}$Si$_{7}$&   0.2 & 216  &  2   & 288 & 5 & 120 & 345816 \ \\
Li$_{7}$Si$_{3}$&   0.1 &  16&  5   &  12 & 5 &  20 &     6008  \ \\
Li$_{13}$Si$_{4}$&   0.2 &  60&  9   &  72 & 5 &  36 &   116700 \ \\
%Li$_{15}$Si$_{4}$&   0.2 &  24&  4   &  60 & 5 &  10 &    12024  \ \\
%Li$_{17}$Si$_{4}$&   0.2 &  64& 14   & 184 & 5 &  42 &   541024  \ \\
Li$_{22}$Si$_{5}$&     4 &  24&  7   & 300 & 5 &  48 &   504480  \ \\
Si               &   0.1 &   4&  0.1 &   2 & 5 &   2 &        4  \ \\ 
\end{tabular}
\label{table_qe_time}
\end{table}

\begin{table}[ht!]
\setlength{\tabcolsep}{1.5pt}
\tiny
\centering
\caption{CPU time using MD for Li$_{x}$Si$_{y}$ crystal structures. The time of a bulk modulus calculation is $t_{B}$, the number of performed NPT calculations is $N_{\text{NPT}}$, 
the number of used cores for each NPT calculations is $N_{\text{cores}}^{\text{NPT}}$, the time of a $\alpha$ calculations is $t_{\alpha}$, the time of a PDOS calculation is $t_{\text{PDOS}}$,
the number of cores used for each PDOS calculation is $N_{\text{cores}}^{\text{PDOS}}$ and the time for one complete MD $C_{p}$ calculation is $t_{\text{MD}}$.\newline}
\centering
\begin{tabular}{l|ccccccc}
System & $t_{B}$ & $N_{\text{NPT}}$ & $N_{\text{cores}}^{\text{NPT}}$ & $t_{\alpha}$ &  $t_{\text{PDOS}}$ & $N_{\text{cores}}^{\text{PDOS}}$ & $t_{\text{MD}}$\ \\ 
                  & [CPUh]  &     &     & [CPUh]  &  [CPUh]   &      & [CPUh]   \ \\ \hline
Li                & 0       & 27  &  40 &  0.12   &   8       &  200 &  11.24      \ \\ 
%LiSi              & 0.01    & 29  &  60 &  0.52   &  47.1     &  400  &  62.19   \ \\
Li$_{12}$Si$_{7}$ & 0.26    & 35  &  22 &  0.17   &   9.77    &   22 &  15.98     \ \\ 
Li$_{7}$Si$_{3}$  & 0       & 15  &  22 &  0.14   &  11.69    &   22 &  13.79    \ \\ 
Li$_{13}$Si$_{4}$ & 0.02    & 29  &  16 &  0.24   &  24.01    &  100 &  31.00   \ \\  
Li$_{22}$Si$_{5}$ & 0.06    & 26  &  40 &  0.57   &  46.56    &  100 &  61.44     \ \\
Si                & 0.05    & 24  &   1 &  0.03   &   3.7     &    1 &   4.47   \ \\  

\end{tabular}
\label{table_lammps_time}
\end{table}

%\FloatBarrier
\section{Conclusion\label{conclusion}} 

% MEAM and DFT comparison 
Within this contribution we showed for a complex material class like lithium silicides it is possible to calculate 
with comparable accuracy to density functional theory mechanical properties such as elastic constants and bulk moduli 
with one MEAM force-field (see TAB. \ref{elastic_properties}).

Based on the fact that both methods deliver suitable elastic constants we were also able to calculate thermodynamic properties based on a phonon density of states with DFT as well as MEAM.
Our proposed MD $C_{p}$ method for the calculation of the isobaric heat capacity turns out to be numerically fast and accurate compared 
to experimental or ab-initio data (cp. TAB. \ref{table_qe_time} and TAB. \ref{table_lammps_time}).\\
% What is great why you should do this ? 
For Li$_{12}$Si$_{7}$ and Li$_{22}$Si$_{5}$, we identified with this method two extended temperature ranges with almost constant but different linear expansion coefficients for both structures. One of this expansion coefficients describes the low temperature region, whereas the other one describes the high temperature region.\\
Furthermore, we present temperature dependent phonon density of states, which in case of Li$_{13}$Si$_{4}$ 
show a better agreement between our MD $C_{p}$ method and the obtained ab-initio data by freezing out the diffusion process using a low temperature phonon density of states. 
% General 
In general the ab-initio thermodynamic calculations are in excellent agreement with experimental data in the low temperature region (0 - 400 K), while the MD $C_{p}$ method based thermodynamic data can deliver 
a faster description in good agreement with experimental data especially in the high temperature limit ($T > 300$ K). Other groups showed that an explicit treatment of anharmonic effects is necessary to overcome the limitations 
of the quasi-harmonic approximation \cite{AnharmonicApproach,AnharmonicApproach2,AnharmonicApproach3}. Our MD $C_{p}$ method based computational thermodynamics is
able to describe the high temperature limit better by using a temperature dependent phonon DOS (see FIG. \ref{Cp_Li13Si4}). Additionally, we are able to evaluate linear expansion coefficients by performing a set of NPT-ensemble MD runs (see FIG. \ref{Cp_Li12Si7}  and FIG. \ref{Cp_Li22Si5}). 
The essential advantage of our MD $C_{p}$ method is that it is capable to describe the structure change due to the effect of temperature (e.g. see radial distribution function (RPDF) FIG. \ref{Cp_Li}~(a) and FIG. \ref{Cp_Li12Si7}~(a)). 
%The essential advantage of our MD $C_{p}$ method is that the structure (e.g. see radial distribution function (RPDF) FIG.(\ref{Cp_Li})~(a) and FIG.(\ref{Cp_Li12Si7})~(a)) is able to change due to the effect of temperature. 
Thus, our method allows to describe diffusion effects as well as the melting behavior of the treated system. 
Based on these essentials of our method, we could observe that Li atoms in the Li$_{22}$Si$_{5}$ phase show amorphous/liquid-like behavior (e.g. RPDF FIG. \ref{Cp_Li22Si5}~(a)).
If this materials behavior is correct it may also explains why it is difficult to distinguish between Li$_{22}$Si$_{5}$, Li$_{21}$Si$_{5}$ and Li$_{17}$Si$_{4}$ experimentally, as all of them have a high and very similar Li content.

\section{Acknowledgements\label{acknowledgements}}
The authors thank the Deutsche Forschungsgemeinschaft DFG (WeNDeLIB - SPP 1473) and the 'Support the Best' program within the Excellence Initiative of the TU Dresden for financial support as well as the ZIH in Dresden for computational time and support.
Furthermore, we especially thank Daniel Thomas for the excellent experimental isobaric heat capacity data, 
which we used as experimental reference. Additionally, we want to thank the institute of theoretical physics and the institute 
of physical chemistry for fruitful discussions and an excellent cooperation. 

%\section*{References}
\bibliography{LiSi}

\end{document}